\documentclass[11pt,a4paper]{article}
\usepackage[T1]{fontenc}
\usepackage{lmodern,microtype}
\usepackage[margin=29mm]{geometry}
\usepackage{amsmath,amssymb,amsthm,mathtools,paralist,xcolor}
\usepackage[UKenglish]{babel}
\definecolor{inkblue}{RGB}{36,83,121}
\definecolor{warmred}{RGB}{166,61,50}
\usepackage[colorlinks=true,linkcolor=inkblue,citecolor=inkblue,urlcolor=inkblue]{hyperref}
\usepackage{thmtools,thm-restate}
\usepackage[capitalise,nameinlink,noabbrev]{cleveref}
\newtheorem{theorem}{Theorem}[section]
\newtheorem{lemma}[theorem]{Lemma}
\newtheorem{proposition}[theorem]{Proposition}

\theoremstyle{definition}
\newtheorem{definition}[theorem]{Definition}
\newtheorem{brule}{Branching Rule}
\crefname{brule}{Branching Rule}{Branching Rules}
\newtheorem{redrule}{Reduction Rule}
\crefname{redrule}{Reduction Rule}{Reduction Rules}
\title{Cluster Editing in Subexponential Fixed-Parameter Time}
\author{Alexander Firbas\thanks{TU Wien, Vienna, Austria. \texttt{alexander.firbas@tuwien.ac.at}}
 \and Manuel Sorge\thanks{TU Wien, Vienna, Austria. \texttt{manuel.sorge@ac.tuwien.ac.at}}}
\date{}
\hypersetup{pdftitle={Cluster Editing in Subexponential Fixed-Parameter Time},pdfauthor={Alexander Firbas, Manuel Sorge}}
\newcommand{\ClusterEditing}{{\normalfont\textsc{Cluster Editing}}}
\newcommand{\WCE}{{\normalfont\textsc{Weighted Cluster Editing}}}
\newcommand{\ThreeSAT}{{\normalfont\textsc{3-SAT}}}
\newcommand{\Dense}{{\normalfont\textsc{Dense}}}
\newcommand{\Sparse}{{\normalfont\textsc{Sparse}}}
\newcommand{\OPT}{\operatorname{OPT}}
\newcommand{\cost}{\operatorname{cost}}
\newcommand{\pos}[1]{\max\{0,#1\}}
\newcommand{\tot}[1]{\lVert #1\rVert}

\newenvironment{lemenum}{\begin{compactenum}[(i)]}{\end{compactenum}}
\begin{document}
\maketitle
\begin{abstract}
\ClusterEditing{} asks whether at most $k$ edge additions and deletions turn a given $n$-vertex graph into a disjoint union of cliques. For more than a decade, the fastest algorithm for it was the golden-ratio algorithm of B{\"o}cker~\cite{boecker-jda} with running time $O(1.62^k+n^{O(1)})$. Cao and Chen~\cite{cao-chen} had conjectured that the subexponential running time $2^{o(k)}\cdot n^{O(1)}$ is possible, but this contradicts the Exponential Time Hypothesis (ETH). Recently, the breakthrough algorithms for \ThreeSAT{} of OpenAI~\cite{openai-eth} and Anthropic~\cite{anthropic-eth} refuted ETH itself.

Our main result transfers running times from \ThreeSAT{} to \ClusterEditing{}. If \ThreeSAT{} on formulas with $m$ clauses can be decided in time $T(m)$, then there is an algorithm that, given a graph with $n$ vertices and integers $k$ and $r$ with $2\le r\le k^{1/3}$, decides \ClusterEditing{} in time
\[
 2^{O((k\log r)/r)}\cdot T\bigl(2^{O(r^2\log r)}\cdot k\bigr)\cdot n^{O(1)}.
\]
Combined with the breakthrough algorithms for \ThreeSAT{}, this proves Cao and Chen's conjecture, as it solves \ClusterEditing{} in time
\[
 2^{O\left(k\cdot(\log\log\log k)^{3/2}/\sqrt{\log\log k}\right)}\cdot n^{O(1)}=2^{o(k)}\cdot n^{O(1)}.
\]
The key ingredient is a sparsification theorem for \ClusterEditing{}, the counterpart of the Sparsification Lemma for satisfiability: In time $2^{O((k\log r)/r)}\cdot n^{O(1)}$, it reduces a given instance to $2^{O((k\log r)/r)}$ weighted instances with $O(rk)$ vertices and degree $O(r)$, each of which is equivalent to a \ThreeSAT{} formula with $2^{O(r^2\log r)}\cdot k$ clauses.
Conversely, by a known reduction with linear blow-up, every saving in the exponent for \ClusterEditing{} is, up to constant factors, a saving for \ThreeSAT{}.
\end{abstract}
\section{Introduction}\label{sec:overview}
\ClusterEditing{} asks whether at most $k$ edge additions and deletions turn a given $n$-vertex graph into a disjoint union of cliques. The problem is NP-hard~\cite{shamir2004}, and for more than a decade the fastest algorithm for it was the golden-ratio algorithm of B{\"o}cker with running time $O(1.62^k+n^{O(1)})$~\cite{boecker-jda}. We show that every algorithm for \ThreeSAT{} yields an algorithm for \ClusterEditing{} whose running time is determined by that of the algorithm for \ThreeSAT{}. With the recent breakthrough algorithms for \ThreeSAT{}~\cite{openai-eth,anthropic-eth}, \ClusterEditing{} can be solved in time $2^{o(k)}\cdot n^{O(1)}$, and hence in time $c^k\cdot n^{O(1)}$ for every $c>1$.

Algorithms with running time $2^{o(k)}\cdot n^{O(1)}$ are called \emph{subexponential} fixed-parameter algorithms.
Pointing to the similarities between \ClusterEditing{} and Feedback Arc Set in Tournaments, which has such an algorithm~\cite[Section~2.3]{edge-mod-survey}, Cao and Chen conjectured that \ClusterEditing{} has one as well~\cite[Section~5]{cao-chen}.
The conjecture was soon refuted assuming \emph{Exponential Time Hypothesis} (ETH).
ETH states that there is no algorithm that decides \ThreeSAT{} on formulas with $v$ variables in time $2^{o(v)}$~\cite{ip2001,ipz2001}.
Here and throughout, algorithms are deterministic.
\begin{theorem}[Komusiewicz and Uhlmann~{\cite[Theorem~2]{komusiewicz-uhlmann2012}}; Fomin et al.~{\cite[Theorem~3]{fomin-pclusters}}]\label{thm:lower}
Unless ETH fails, \ClusterEditing{} cannot be solved in time $2^{o(k)}\cdot n^{O(1)}$. This holds even for graphs with maximum degree six.
\end{theorem}

However, ETH itself has recently been refuted.
We state the refutation in the form in which we use it, with the size of a formula measured by its number of clauses.
\begin{theorem}[Anthropic~{\cite[Theorem~1]{anthropic-eth}}]\label{thm:sat}
  There are an integer $a\ge1$ and an algorithm that decides whether a given 3-CNF formula with $m\ge2$ clauses is satisfiable in time at most $2^{a\cdot m/\sqrt{\log m}}$.
\end{theorem}
A refutation of ETH was announced independently by OpenAI~\cite[Theorem~1.1]{openai-eth}.

With ETH gone, \cref{thm:lower} the conjecture of Cao and Chen was open again but it does not directly follow from a subexponential algorithm for \ThreeSAT{}: A subexponential algorithm for \ThreeSAT{} transfers to another problem in particular through reductions to that produces formulas of size linear in the parameter.
Such reductions are known, for instance, for problems whose solution is a set of vertices, which one variable per vertex describes~\cite{ipz2001,ganian-emso}.
However, a solution of \ClusterEditing{} is a partition of the vertex set into arbitrarily many clusters.
Even after kernelization to $O(k)$ vertices~\cite{guo-kernel,cao-chen}, writing down a partition takes on the order of $k\log k$ bits, and a formula with $k\log k$ variables is of no use: the time $2^{o(k\log k)}$ that it yields for \ClusterEditing\ is not better than the running times of the standard branching algorithms.

\paragraph{Our results.}
Our main result transfers a running time for \ThreeSAT{} to \ClusterEditing{} in the following way. We say that \ThreeSAT{} can be decided in time $T(m)$, for a nondecreasing function $T\ge1$, if some algorithm decides every 3-CNF formula with $m$ clauses in time at most $T(m)$ (\cref{sec:model}).
\begin{restatable}[Transfer]{theorem}{transferthm}\label{thm:main}
There are integers $c_2,c_3,\dots$ with $c_r=2^{O(r^2\log r)}$ such that the following holds. Let $T$ be a nondecreasing function such that \ThreeSAT{} can be decided in time $T(m)$. Then there is an algorithm that, given a graph $G$ with $n$ vertices and integers $k$ and $r$ with $2\le r\le k^{1/3}$, decides whether $(G,k)$ is a yes-instance of \ClusterEditing{} in time
\[
 2^{O((k\log r)/r)}\cdot T(c_r\cdot k)\cdot n^{O(1)}.
\]
The constants hidden in the $O$-notation depend neither on $r$ nor on $T$.
\end{restatable}
The integer $r$ trades two costs against each other. Our algorithm calls the algorithm for \ThreeSAT{} $2^{O((k\log r)/r)}$ times, which is the fewer the larger $r$ is, and each call is for a formula with at most $c_r\cdot k$ clauses, which is the shorter the smaller $r$ is; the number $c_r$ is defined in \cref{eq:cr}. Since the constants do not depend on $r$, we may let $r$ grow with $k$. With the bound of \cref{thm:sat} and $r=\lfloor\sqrt{(\log\log k)/(C\log\log\log k)}\rfloor$ for a suitable constant $C$, we obtain our second result, which proves the conjecture of Cao and Chen.
\begin{restatable}{corollary}{maincor}\label{cor:main}
\ClusterEditing{} can be solved in time $2^{O(k/s(k))}\cdot n^{O(1)}$, where
\[
 s(k):=\frac{\sqrt{\log\log k}}{(\log\log\log k)^{3/2}}\quad\text{for }k\ge2^{16}
\]
and $s(k):=1$ for $k<2^{16}$. In particular, \ClusterEditing{} can be solved in time $2^{o(k)}\cdot n^{O(1)}$, and in time $c^k\cdot n^{O(1)}$ for every $c>1$.
\end{restatable}
The saving $s(k)$ grows slowly, for two reasons: the saving $\sqrt{\log m}$ of \cref{thm:sat} is small itself, and the growth of $c_r$ keeps $r$ small (\cref{sec:conclusion}). Every improvement of \cref{thm:sat} improves \cref{cor:main} through \cref{thm:main}. \Cref{cor:main} is an asymptotic statement and does not compete with the branching algorithms on instances of practical size: already $c_2>2^{6000}$.

\paragraph{Sparsification.}
The first half of the proof of \cref{thm:main} is a statement about \ClusterEditing{} alone and can be stated as follows.
It concerns the weighted variant of the problem in which every vertex pair has an integer weight, positive if the pair should be joined and negative if it should be separated; we write $\OPT(W)$ for the minimum cost of a partition of a weighted instance $W$ and call $(W,k)$ \emph{$r$-sparse} if all weights of $W$ are odd, every vertex is positive to fewer than $4r$ others, $W$ has at most $16rk$ vertices, and no connected component of the graph formed by the positive pairs is complete (\cref{sec:model,def:sparse}).
\begin{restatable}[Sparsification]{theorem}{sparsificationthm}\label{thm:sparsification}
There is an algorithm that, given a graph $G$ with $n$ vertices and integers $k$ and $r$ with $2\le r\le\sqrt k$, runs in time $2^{O((k\log r)/r)}\cdot n^{O(1)}$ and outputs at most $2^{(4k\log r)/r}$ $r$-sparse instances $(W_i,k_i)$ with $k_i\le k$ and total weight at most $n^2$ such that $(G,k)$ is a yes-instance of \ClusterEditing{} if and only if $\OPT(W_i)\le k_i$ for some $i$. The constants hidden in the $O$-notation do not depend on $r$.
\end{restatable}
For every $\varepsilon>0$ and a suitable constant $r$, the theorem replaces an instance of \ClusterEditing{} by at most $2^{\varepsilon k}$ instances with $O(k)$ vertices and $O(k)$ positive pairs.
This is the role that the Sparsification Lemma of Impagliazzo, Paturi, and Zane~\cite{ipz2001} plays for \ThreeSAT{}, with the difference that our output instances are weighted.
The bound on the number of positive neighbours make it possible to model the resulting instances with a short \ThreeSAT{} formula: in an $r$-sparse instance, a cluster of bounded size that contains a given vertex is one of a bounded number of connected sets, so that a partition becomes a set of such sets, chosen from a family of linear size.

\paragraph{Tightness.}
\ClusterEditing{} is at least as hard as \ThreeSAT{}. Fomin et al.\ map a 3-CNF formula with $m$ clauses in polynomial time to an equivalent instance of \ClusterEditing{} with $O(m)$ vertices and $k=O(m)$~\cite[Lemma~18]{fomin-pclusters}. Hence, if $f$ and $k\mapsto k/f(k)$ are nondecreasing functions and \ClusterEditing{} can be solved in time $2^{O(k/f(k))}\cdot n^{O(1)}$, then \ThreeSAT{} can be decided in time $2^{O(m/f(m))}\cdot m^{O(1)}$. So the saving $s(k)$ of \cref{cor:main} cannot be improved beyond the best saving for \ThreeSAT{} without improving the latter. In the other direction, \cref{thm:main} turns the saving $\sqrt{\log m}$ of \cref{thm:sat} into the smaller $s(k)$, and we do not know whether this loss is necessary.

\paragraph{Outline of the proof.}
We work with the weighted variant throughout; a positive weight is the price of separating the pair, and the absolute value of a negative weight is the price of joining it. The proof has two phases. Phase one is a search tree that leads to instances in which positively related vertices have similar neighbourhoods. Phase two shows that such instances fall apart into dense parts, which are solved directly, and sparse parts, which are encoded into formulas.

\paragraph{Phase one: branching down to so-called $r$-uniform instances (\cref{sec:model,sec:branching}).}
On the weighted variant there is a well-known two-way branch on a positive pair $\{u,v\}$: either pay for separating the pair, or merge $u$ and $v$ into one vertex. A merge is not free: a third vertex that is positive to one of $u,v$ and negative to the other must be in conflict with one of them once they share a cluster. We call the price forced in this way the \emph{merge cost} of the pair. We branch only on pairs whose merge cost is at least the threshold $r$. Then the two children lower the budget by at least $1$ and at least $r$, and the corresponding branching number tends to $1$ as $r$ grows: the search tree has at most $r^{4k/r}=2^{(4k\log r)/r}$ leaves (\cref{prop:phase-one}). At a leaf, every positive pair has merge cost less than $r$; we call such instances \emph{$r$-uniform} (\cref{def:uniform}). In an $r$-uniform instance, any two vertices that are joined by a positive pair have the same positive neighbours up to fewer than $r$ exceptions.

The technical difficulty in this phase is that merging creates pairs of weight zero. A vertex that has weight zero to $u$ adds nothing to the merge cost of $\{u,v\}$, whatever its weight to $v$ is. Small merge costs then no longer imply that $u$ and $v$ have similar positive neighbourhoods, and this implication is all that phase two knows about its input. We overcome the difficulty by a parity argument. Initially all weights are $\pm1$, and we maintain that all weights are odd, hence nonzero. Merging two vertices breaks this property only at the merged vertex $z$, whose weights are sums of two odd numbers and thus even. We repair it at once by a second branching rule, which takes a positive neighbour $v$ of $z$ and either merges $z$ and $v$ or pays for separating them. The merge adds the odd weights of $v$ to the even weights of $z$, so all weights are odd again. The separation costs at least $2$, because the weight of $\{z,v\}$ is positive and even, and it leaves $z$ with one positive neighbour fewer. Once $z$ has no positive neighbour, it forms a cluster by itself at no cost, and we delete it. The second rule creates a difficulty of its own for the analysis: its merge may cost nothing, so the rule alone does not have a bounded branching number. However, a merge of the second rule always returns to an instance with only odd weights, and from there the next vertex with even weights is at least $r$ units of budget away. We therefore bound the search tree by two mutually dependent recurrences, one for instances with only odd weights and one for instances with a merged vertex.

\paragraph{Phase two, first step: from $r$-uniform to $r$-sparse instances (\cref{sec:dichotomy,sec:dense,sec:to-sparse}).}
Consider the graph whose edges are the positive pairs of an $r$-uniform instance. Its connected components can be solved independently, and the similarity of neighbourhoods forces a dichotomy on each of them: either every vertex has fewer than $4r$ neighbours, or every vertex has fewer than $4r$ non-neighbours (\cref{lem:dichotomy}). We call the component sparse in the first case and dense in the second.

In a dense component, a partition with more than one cluster has a cluster $C$ with at most half of the vertices, and each vertex of $C$ is positive to almost half of all vertices outside its cluster. Splitting off a cluster is therefore so expensive that we can guess the clusters one by one. This takes time $2^{O(\sqrt k\log k)}\cdot n^{O(1)}$ if $r\le\sqrt k$ (\cref{prop:dense}), so we can compute the optimum of every dense component and subtract it from the budget.

The sparse components that are not cliques have $O(rk)$ vertices in total, unless they exceed the budget: few large clusters contain many negative pairs, and many small clusters separate many positive pairs (\cref{lem:sparse-size}). This bound takes the role that a problem kernel would otherwise play. What remains is an $r$-sparse instance (\cref{prop:to-sparse}), which completes the proof of \cref{thm:sparsification}.

\paragraph{Phase two, second step: from $r$-sparse instances to formulas (\cref{sec:sparse}).}
Two facts turn an $r$-sparse instance with budget $k$ into formulas with $c_r\cdot k$ clauses. First, clusters may be assumed to be connected in the graph of positive pairs, and in a graph in which every vertex has fewer than $4r$ neighbours, the number of connected sets of a given size that contain a given vertex is bounded in terms of $r$ and that size. With one variable for each such set, choosing a partition becomes an exact cover problem with a number of variables that is linear in the number of vertices. Second, the budget is a threshold on the sum of the costs of the chosen sets, and a circuit with linearly many gates checks it, provided that each set has bounded cost. The clusters that are too large or too expensive for this encoding contain few vertices in total; we guess them, which multiplies the running time by $2^{k/r}$ (\cref{prop:sparse}).

\paragraph{Choosing the threshold (\cref{sec:proof}).}
The search tree has $2^{(4k\log r)/r}$ leaves, each leaf leads to $2^{k/r}$ formulas, and each formula has at most $c_r\cdot k$ clauses; this gives \cref{thm:main}. For \cref{cor:main} we choose $r$ as large as the bound of \cref{thm:sat} permits, namely such that $c_r$ is at most $(\log k)^{1/4}$.

 \section{Preliminaries}\label{sec:model}
Merging two vertices, the central operation of phase one (\cref{sec:overview}), does not stay within unweighted graphs: the merged vertex relates to a third vertex by the sum of two relations. This section therefore fixes the weighted problem on which all later sections operate, shows how \ClusterEditing{} embeds into it, and singles out the instances with only odd weights, among which are the so-called $r$-uniform instances that phase two solves (\cref{def:uniform}).

\paragraph{Basic notation.}
All graphs are finite, undirected, and have no loops or parallel edges; we write $\{u,v\}$ for an undirected edge and, more generally, for an unordered pair of distinct vertices. For a graph $G$, we write $V(G)$ for its vertex set, $G[C]$ for the subgraph induced by a set $C\subseteq V(G)$, $N(u)$ for the set of neighbours of a vertex $u$, and $N[u]:=N(u)\cup\{u\}$. The \emph{degree} of $u$ is $|N(u)|$, and $\Delta(G)$ is the maximum degree of a vertex of $G$. The \emph{complement} $\overline G$ of $G$ has the same vertex set, and two distinct vertices are adjacent in $\overline G$ if and only if they are not adjacent in $G$; thus $\Delta(\overline G)$ is the maximum number of non-neighbours that a vertex has in $G$. For a real number $a$ we write $|a|$ for its absolute value; note that $\pos a=(|a|+a)/2$. Empty sums are zero. We write $\log$ for the binary logarithm, $\ln$ for the natural logarithm, and $e$ for Euler's number.

\paragraph{Formulas.}
A \emph{literal} is a Boolean variable or its negation, a \emph{clause} is a disjunction of literals, and a \emph{3-CNF formula} is a conjunction of clauses with at most three literals each. \ThreeSAT{} is the problem of deciding whether a given 3-CNF formula has a satisfying assignment. An \emph{assignment} gives each variable one of the values true and false, which we identify with $1$ and $0$; it \emph{satisfies} a clause if it makes at least one of its literals true, and it satisfies a formula if it satisfies all its clauses. A clause without literals, the \emph{empty clause}, is satisfied by no assignment.

We measure a 3-CNF formula by its number of clauses; the variables of a formula are those that occur in its clauses, so a formula with $m$ clauses has at most $3m$ variables. Throughout, $T$ denotes a nondecreasing function from the nonnegative integers to the real numbers with $T(m)\ge1$ for every $m$. We say that \emph{\ThreeSAT{} can be decided in time $T(m)$} if there is an algorithm that, for every $m\ge0$, decides whether a given 3-CNF formula with $m$ clauses is satisfiable in time at most $T(m)$. Since $T$ is nondecreasing, such an algorithm takes time at most $T(m)$ on every formula with at most $m$ clauses. As everywhere in this paper, algorithms are deterministic.

\paragraph{Weighted instances.}
A \emph{weighted instance} is a symmetric integer matrix $W=(w_{uv})_{u,v\in V}$ with zero diagonal, indexed by a finite set $V$ of \emph{vertices}; we write $n:=|V|$. A \emph{partition} $\mathcal{P}$ of $V$ is a family of pairwise disjoint nonempty subsets of $V$, called \emph{clusters}, whose union is $V$. A partition \emph{joins} a pair $\{u,v\}$ if $u$ and $v$ lie in the same cluster, and \emph{separates} the pair otherwise. The \emph{cost} of $\mathcal{P}$ is
\begin{align}
 \cost_W(\mathcal{P}):=
 \sum_{\{u,v\}\text{ separated by }\mathcal{P}}\pos{w_{uv}}
 +\sum_{\{u,v\}\text{ joined by }\mathcal{P}}\pos{-w_{uv}},\label{eq:cost}
\end{align}
where both sums run over unordered pairs of distinct vertices. That is, separating a pair of positive weight costs that weight, joining a pair of negative weight costs the absolute value of that weight, and a pair of weight zero costs nothing either way. Every pair has nonnegative cost, so $\cost_W(\mathcal{P})\ge0$. Define $\OPT(W):=\min_\mathcal{P}\cost_W(\mathcal{P})$, the minimum being over all partitions of $V$; the empty instance has the empty partition as its only partition and $\OPT(W)=0$.

\WCE{} is the following decision problem: given a weighted instance $W$ and an integer $k$, the \emph{budget}, decide whether $\OPT(W)\le k$. We call the pair $(W,k)$ a \emph{yes-instance} if $\OPT(W)\le k$ and a \emph{no-instance} otherwise. Since costs are nonnegative, every pair $(W,k)$ with $k<0$ is a no-instance.

A graph $G$ with vertex set $V$ gives the weighted instance $W_G$ with weight $+1$ on every edge and $-1$ on every nonedge. A partition $\mathcal{P}$ of $V$ corresponds to the disjoint union of cliques on its clusters, and $\cost_{W_G}(\mathcal{P})$ counts the edges of $G$ between clusters plus the nonedges of $G$ inside clusters, that is, the edge deletions and additions that turn $G$ into this union of cliques. Hence $(G,k)$ is a yes-instance of \ClusterEditing{} if and only if $(W_G,k)$ is a yes-instance of \WCE{}.

\paragraph{Positive pairs and odd weights.}
A pair $\{u,v\}$ is \emph{positive} if $w_{uv}>0$ and \emph{negative} if $w_{uv}<0$; we then also say that $u$ is \emph{positive to} $v$, respectively \emph{negative to} $v$. A \emph{positive neighbour} of $u$ is a vertex that is positive to $u$. A weighted instance is \emph{all-odd} if all its off-diagonal weights are odd. In an all-odd instance no weight is zero, so every pair is either positive or negative, and every positive or negative pair costs at least one when the partition treats it against its sign. The instance $W_G$ is all-odd.

\paragraph{Total weight and running times.}
The \emph{total weight} of $W$ is $\tot W:=\sum_{\{u,v\}}|w_{uv}|$, the sum being over unordered pairs of distinct vertices. Every weight has absolute value at most $\tot W$, and $\tot{W_G}\le n^2$. None of our operations increases the total weight (\cref{lem:safe}), so all numbers that arise from an input $W_G$ stay polynomially bounded in $n$. We state running times in terms of $n$, $k$, and $\tot W$.
 \section{Phase one: branching on pairs with large merge cost}\label{sec:branching}
This section reduces \WCE{} on all-odd instances (\cref{sec:model}) to so-called $r$-uniform instances (\cref{def:uniform}), at the price of a search tree with at most $r^{4k/r}$ leaves, where $k$ is the budget and the integer $r\ge2$ is a parameter that we are free to choose; the base $r^{4/r}$ of this bound tends to $1$ as $r$ grows. \Cref{sec:flip-merge} defines the two operations that decide whether a positive pair is separated or joined, and \cref{sec:rules} builds two branching rules and one reduction rule from them. \Cref{sec:tree} bounds the size of the resulting search tree by two mutually dependent recurrences. \Cref{sec:uniform} defines $r$-uniform instances, which are the all-odd instances to which no rule applies, and states the reduction (\cref{prop:phase-one}).

\subsection{Flipping and merging}\label{sec:flip-merge}
To branch on a positive pair $\{u,v\}$ of a weighted instance $W$ we need one instance that represents the partitions separating the pair and one that represents the partitions joining it. Both are standard for weighted cluster editing~\cite{boecker-weighted,boecker-jda}, except that we do not forbid a separated pair from being joined later; we only make joining it expensive. This keeps all weights finite and, more importantly for us, preserves their parity.

Let $\{u,v\}$ be a positive pair of $W$. The \emph{flip} $W^{\{u,v\}}$ is the weighted instance obtained from $W$ by changing the weight of $\{u,v\}$ from $w_{uv}$ to $-w_{uv}$ and nothing else. The following lemma states that, after paying $w_{uv}$, the flip has the same costs as $W$ on partitions that separate the pair and larger costs on all others.
\begin{lemma}[Flip]\label{lem:flip}
Let $\{u,v\}$ be a positive pair of a weighted instance $W$ and let $\mathcal{P}$ be a partition of its vertex set. If $\mathcal{P}$ separates $\{u,v\}$, then $\cost_W(\mathcal{P})=w_{uv}+\cost_{W^{\{u,v\}}}(\mathcal{P})$. If $\mathcal{P}$ joins $\{u,v\}$, then $\cost_W(\mathcal{P})=\cost_{W^{\{u,v\}}}(\mathcal{P})-w_{uv}$.
\end{lemma}
\begin{proof}
The two instances differ only in the weight of $\{u,v\}$, so by \cref{eq:cost} the two costs differ only in the summand of this pair. If $\mathcal{P}$ separates the pair, the summand is $\pos{w_{uv}}=w_{uv}$ under $W$ and $\pos{-w_{uv}}=0$ under the flip. If $\mathcal{P}$ joins the pair, the summand is $\pos{-w_{uv}}=0$ under $W$ and $\pos{w_{uv}}=w_{uv}$ under the flip.
\end{proof}

The \emph{merge} $W/\{u,v\}$ is the weighted instance obtained from $W$ by replacing $u$ and $v$ with one new vertex $z$ and setting $w_{zy}:=w_{uy}+w_{vy}$ for every other vertex $y$; weights between other vertices are unchanged. If a third vertex $y$ is positive to one of $u,v$ and negative to the other, then every partition that joins $u$ and $v$ pays for $y$: it pays the positive weight if it separates $y$ from the two and the absolute value of the negative weight if it joins $y$ with them. In the sum $w_{zy}$ the smaller of the two absolute values has cancelled. The \emph{merge cost} of the positive pair $\{u,v\}$ collects these cancelled amounts:
\begin{align}
 \mu_W(u,v):=\sum_{y\in V\setminus\{u,v\}}\frac{|w_{uy}|+|w_{vy}|-|w_{uy}+w_{vy}|}{2}.\label{eq:merge-cost}
\end{align}
The summand of $y$ is zero if $w_{uy}$ and $w_{vy}$ have the same sign or one of them is zero, and it equals $\min\{|w_{uy}|,|w_{vy}|\}$ if they have opposite signs. In particular $\mu_W(u,v)$ is a nonnegative integer. For example, if $w_{uy}=3$ and $w_{vy}=-1$, then $w_{zy}=2$ and the summand of $y$ is $1$: separating $y$ from $u$ and $v$ costs $3=2+1$ under $W$, and joining it with them costs $1=0+1$.
\begin{lemma}[Merge]\label{lem:merge}
Let $\{u,v\}$ be a positive pair of a weighted instance $W$ and let $z$ be the new vertex of $W/\{u,v\}$. Replacing $u$ and $v$ by $z$ in their common cluster is a bijection from the partitions of $W$ that join $\{u,v\}$ to the partitions of $W/\{u,v\}$. If it maps $\mathcal{P}$ to $\mathcal{P}'$, then $\cost_W(\mathcal{P})=\mu_W(u,v)+\cost_{W/\{u,v\}}(\mathcal{P}')$.
\end{lemma}
\begin{proof}
The map is a bijection because its inverse replaces $z$ by $u$ and $v$. We compare the two costs pair by pair. The pair $\{u,v\}$ is positive and joined by $\mathcal{P}$, so it costs nothing under $W$, and it is absent from $W/\{u,v\}$. A pair that avoids $u$, $v$, and $z$ has the same weight in both instances and is joined by $\mathcal{P}$ if and only if it is joined by $\mathcal{P}'$. It remains to compare, for each vertex $y\notin\{u,v\}$, the cost of the two pairs $\{u,y\}$ and $\{v,y\}$ under $W$ with the cost of $\{z,y\}$ under $W/\{u,v\}$. Put $a:=w_{uy}$ and $c:=w_{vy}$, and recall that $\pos x=(|x|+x)/2$. If $\mathcal{P}$ separates $y$ from $u$ and $v$, the difference of these costs is
\[
 \pos a+\pos c-\pos{a+c}=\frac{|a|+|c|-|a+c|}{2}.
\]
If $\mathcal{P}$ joins $y$ with $u$ and $v$, the difference is $\pos{-a}+\pos{-c}-\pos{-(a+c)}$, which is the same expression. In both cases it is the summand of $y$ in \cref{eq:merge-cost}. Summing over all $y$ gives the claim.
\end{proof}

\subsection{The three rules}\label{sec:rules}
\Cref{lem:flip,lem:merge} yield a two-way branch on any positive pair. We restrict it in two ways. First, we want all-odd instances (\cref{sec:model}) at the leaves of the search tree, since phase two requires its input to be all-odd; but merging two vertices of an all-odd instance gives the new vertex only even weights, each being a sum of two odd numbers. Second, the branch must make progress in the budget, and a merge may cost nothing. The first branching rule therefore applies only to pairs with merge cost at least a threshold, and a second branching rule and a reduction rule dispose of the vertex with even weights.

The rules create even weights at only one vertex at a time, so the following notion describes all instances that are not all-odd. A \emph{marked instance} $(W,z)$ consists of a weighted instance $W$ and a vertex $z$ of $W$, the \emph{marked vertex}, such that $w_{zy}$ is even for every vertex $y\ne z$ and $w_{xy}$ is odd for all distinct vertices $x,y$ other than $z$. For example, the merge of a positive pair of an all-odd instance, together with its new vertex, is a marked instance. Throughout, $r\ge2$ is a fixed integer, the \emph{threshold}. It is a parameter of the whole algorithm: a larger threshold gives a smaller search tree (\cref{sec:tree}) but leaves more work to phase two. Each branching rule replaces an instance with budget by two instances with budgets, and the reduction rule replaces it by one.

\begin{brule}[Main branch]\label{rule:main}
Let $W$ be an all-odd instance with budget $k$, and let $\{u,v\}$ be a positive pair with $\mu_W(u,v)\ge r$. Branch into
\begin{compactitem}
\item the flip $W^{\{u,v\}}$ with budget $k-w_{uv}$, and
\item the merge $W/\{u,v\}$, marked at its new vertex, with budget $k-\mu_W(u,v)$.
\end{compactitem}
\end{brule}

\begin{brule}[Finishing branch]\label{rule:finish}
Let $(W,z)$ be a marked instance with budget $k$, and let $v$ be a positive neighbour of $z$. Branch into
\begin{compactitem}
\item the flip $W^{\{z,v\}}$, marked at $z$, with budget $k-w_{zv}$, and
\item the merge $W/\{z,v\}$ with budget $k-\mu_W(z,v)$.
\end{compactitem}
\end{brule}

\begin{redrule}[Isolation]\label{rule:isolate}
Let $(W,z)$ be a marked instance with budget $k$ such that $z$ has no positive neighbour. Delete $z$, that is, replace $W$ by its submatrix indexed by $V\setminus\{z\}$, and keep the budget $k$.
\end{redrule}

A rule is \emph{safe} if the instance it is applied to is a yes-instance if and only if at least one of the instances it produces is a yes-instance. The next lemma states that the three rules are safe and records how they change the kind of instance, the budget, and the number of vertices. The analysis in \cref{sec:tree} uses only these facts.
\begin{lemma}[Safety and progress]\label{lem:safe}
\Cref{rule:main,rule:finish,rule:isolate} are safe, and no instance they produce has more vertices or larger total weight than the instance they are applied to. Moreover:
\begin{lemenum}
\item\label{it:main} In \cref{rule:main}, the flip is all-odd and has budget at most $k-1$. The merge is a marked instance, has budget at most $k-r$, and has one vertex fewer.
\item\label{it:finish} In \cref{rule:finish}, the flip is a marked instance and has budget at most $k-2$. The merge is all-odd, has budget at most $k$, and has one vertex fewer.
\item\label{it:isolate} In \cref{rule:isolate}, the produced instance is all-odd, has budget $k$, and has one vertex fewer.
\end{lemenum}
\end{lemma}
\begin{proof}
\emph{Safety of the two branches.} Let $\{u,v\}$ be any positive pair of a weighted instance $W$ with budget $k$; this covers \cref{rule:main,rule:finish}. Suppose first that $\cost_W(\mathcal{P})\le k$ for some partition $\mathcal{P}$. If $\mathcal{P}$ separates $\{u,v\}$, then $\cost_{W^{\{u,v\}}}(\mathcal{P})=\cost_W(\mathcal{P})-w_{uv}\le k-w_{uv}$ by \cref{lem:flip}, so the flip is a yes-instance. If $\mathcal{P}$ joins $\{u,v\}$, then the corresponding partition of the merge costs $\cost_W(\mathcal{P})-\mu_W(u,v)\le k-\mu_W(u,v)$ by \cref{lem:merge}, so the merge is a yes-instance. Conversely, if a partition $\mathcal{P}$ costs at most $k-w_{uv}$ under the flip, then by \cref{lem:flip} it costs at most $(k-w_{uv})+w_{uv}=k$ under $W$, whether or not it separates the pair. If a partition costs at most $k-\mu_W(u,v)$ under the merge, then by \cref{lem:merge} the corresponding partition costs at most $k$ under $W$.

\emph{Safety of the isolation.} Let $W'$ be the instance obtained by deleting $z$. We show $\OPT(W)=\OPT(W')$. Removing $z$ from its cluster in a partition of $W$, and dropping that cluster if it becomes empty, gives a partition of $W'$; it removes the pairs containing $z$ from \cref{eq:cost} and leaves all other summands unchanged, so the cost does not increase. Conversely, adding the cluster $\{z\}$ to a partition of $W'$ gives a partition of $W$ that separates every pair containing $z$. These pairs have weight at most zero because $z$ has no positive neighbour, so they cost nothing, and the cost is unchanged.

\emph{Vertices and total weight.} A flip changes neither the vertex set nor any absolute value of a weight. A merge replaces two vertices by one, and $|w_{uy}+w_{vy}|\le|w_{uy}|+|w_{vy}|$ shows that the total weight does not increase. Deleting a vertex removes one vertex and nonnegative summands of the total weight.

\emph{Kinds of instances and budgets.} For \ref{it:main}, the flip replaces the odd weight $w_{uv}$ by the odd weight $-w_{uv}$, so it is all-odd, and $w_{uv}\ge1$. In the merge, every weight at the new vertex is a sum of two odd numbers and thus even, and all other weights are odd as in $W$; the budget is $k-\mu_W(u,v)\le k-r$. For \ref{it:finish}, the flip changes only the sign of the weight $w_{zv}$, which is even, so the instance remains marked at $z$; since $w_{zv}$ is positive and even, it is at least $2$. In the merge, every weight at the new vertex is the sum of an even weight at $z$ and an odd weight at $v$, hence odd; all other weights are between vertices other than $z$ and are odd. The budget is at most $k$ because merge costs are nonnegative. For \ref{it:isolate}, all weights between vertices other than $z$ are odd.
\end{proof}

\subsection{The size of the search tree}\label{sec:tree}
We now apply \cref{rule:main,rule:finish,rule:isolate} exhaustively and bound the size of the resulting search tree in terms of the budget alone. Measured by the budget, the main branch has branching vector $(1,r)$, whose branching number tends to $1$ as $r$ grows. The finishing branch, however, has branching vector $(2,0)$: by \cref{lem:safe}\ref{it:finish} its merge may leave the budget unchanged. This does no harm, because that merge is an all-odd instance, and the only way from an all-odd instance back to a marked instance is a merge of the main branch, which lowers the budget by at least $r$. We make this precise by bounding the number of leaves below all-odd instances and below marked instances by two different functions of the budget.

The two functions are $b_r^t$ and $a_r\cdot b_r^t$, where $t$ is the budget and the numbers $b_r$ and $a_r$ depend only on the threshold $r\ge2$:
\begin{align}
 b_r:=r^{4/r}=2^{(4\log r)/r}\qquad\text{and}\qquad a_r:=\frac{1}{1-b_r^{-2}}.\label{eq:base}
\end{align}
For example, $b_2=4$, $b_{16}=2$, and $b_{256}=2^{1/8}<1.1$. The following inequality is what the recurrence for all-odd instances needs; its two summands belong to the two children of the main branch.
\begin{lemma}[Base of the search tree]\label{lem:base}
For every integer $r\ge2$ we have $b_r>1$, $a_r\ge1$, and
\begin{align}
 b_r^{-1}+a_r\cdot b_r^{-r}\le1.\label{eq:threshold}
\end{align}
\end{lemma}
\begin{proof}
Since $r>1$, we have $b_r>1$, hence $0<b_r^{-2}<1$ and $a_r\ge1$. For \cref{eq:threshold} we first bound $b_r$ from below. By $e^x\ge1+x$ and $\ln2>1/2$, which holds because $e<4$,
\[
 b_r=e^{(4\ln r)/r}\ge1+\frac{4\ln r}{r}\ge1+\frac{4\ln 2}{r}>1+\frac2r=\frac{r+2}{r}.
\]
Hence $b_r^{-1}<r/(r+2)$ and therefore $1-b_r^{-1}>2/(r+2)\ge1/r$, where the last step uses $r\ge2$. Since $b_r^{-2}<b_r^{-1}$, also $1-b_r^{-2}>1/r$, that is, $a_r<r$. Finally $b_r^{-r}=r^{-4}$, so
\[
 a_r\cdot b_r^{-r}<r\cdot r^{-4}=r^{-3}\le\frac1r<1-b_r^{-1}.\qedhere
\]
\end{proof}

The \emph{search tree} with threshold $r$ of an all-odd instance $W$ with budget $k\ge0$ is the following rooted tree whose nodes are all-odd or marked instances with nonnegative budgets. The root is $W$ with budget $k$. At an all-odd node, if some positive pair has merge cost at least $r$, apply \cref{rule:main} to one such pair. At a marked node, apply \cref{rule:finish} to some positive neighbour of the marked vertex if there is one, and \cref{rule:isolate} otherwise. The children of the node are the produced instances whose budget is nonnegative; produced instances with negative budget are discarded. A \emph{leaf} is a node without children. Thus a leaf is either an all-odd instance in which every positive pair has merge cost less than $r$, or a node at which a branching rule was applied and both produced instances were discarded. The tree is finite by the following lemma.
\begin{lemma}[Search-tree size]\label{lem:tree}
Let $r\ge2$ be an integer, and consider a search tree with threshold $r$ whose root has $n$ vertices and budget $k\ge0$. Recall $b_r$ and $a_r$ from \cref{eq:base}.
\begin{lemenum}
\item\label{it:depth} Every path from the root to a leaf has at most $k+n$ edges.
\item\label{it:leaves} For every node with budget $t$, the number of leaves in the subtree rooted at this node is at most $b_r^t$ if the node is all-odd and at most $a_r\cdot b_r^t$ if it is marked. In particular, the tree has at most $b_r^k$ leaves and at most $(k+n+1)\cdot b_r^k$ nodes.
\end{lemenum}
\end{lemma}
\begin{proof}
Write $b:=b_r$ and $c:=a_r$; then $b>1$ and $c\ge1$ by \cref{lem:base}.

\ref{it:depth} By \cref{lem:safe}, along every edge of the tree neither the budget nor the number of vertices increases, and at least one of them decreases by at least one: the budget for the two flips, and the number of vertices for the two merges and the isolation. Budgets of nodes are nonnegative integers at most $k$, and numbers of vertices are nonnegative integers at most $n$. Hence a path from the root has at most $k$ edges along which the budget decreases and at most $n$ further edges.

\ref{it:leaves} By \ref{it:depth} the tree is finite, so we may use induction on the height of the node, that is, on the maximum number of edges on a path from the node to a leaf below it. Let $t\ge0$ be the budget of the node. Since $b>1$ and $c\ge1$, we have $1\le b^t\le c\cdot b^t$; this settles the case that the node is a leaf. Otherwise the number of leaves below the node is the sum of the numbers of leaves below its children, and the induction hypothesis applies to each child. We bound this sum by adding the bound for every produced instance, whether or not it was discarded; the budget bounds are those of \cref{lem:safe}.

\emph{The node is all-odd.} Then \cref{rule:main} was applied. The flip is all-odd with budget at most $t-1$, and the merge is marked with budget at most $t-r$. The number of leaves is at most
\[
 b^{t-1}+c\cdot b^{t-r}=b^t\cdot\bigl(b^{-1}+c\cdot b^{-r}\bigr)\le b^t
\]
by \cref{eq:threshold} of \cref{lem:base}.

\emph{The node is marked and \cref{rule:finish} was applied.} The flip is marked with budget at most $t-2$, and the merge is all-odd with budget at most $t$. The number of leaves is at most
\[
 c\cdot b^{t-2}+b^t=b^t\cdot\bigl(1+c\cdot b^{-2}\bigr)=c\cdot b^t,
\]
where the last equality is the definition of $a_r$ in \cref{eq:base}, rewritten as $c\cdot(1-b^{-2})=1$.

\emph{The node is marked and \cref{rule:isolate} was applied.} Its only child is all-odd with budget $t$, so the number of leaves is at most $b^t\le c\cdot b^t$.

Finally, the root is all-odd with budget $k$, so there are at most $b^k$ leaves. Every node lies on a path from the root to a leaf, and each such path has at most $k+n+1$ nodes by \ref{it:depth}.
\end{proof}

\subsection{\texorpdfstring{$r$}{r}-uniform instances}\label{sec:uniform}
The leaves of the search tree of \cref{sec:tree} that are not trivially no-instances are all-odd instances to which the main branch (\cref{rule:main}) does not apply. We call these instances $r$-uniform; phase two solves them.
\begin{definition}\label{def:uniform}
Let $r\ge1$ be an integer. A weighted instance is \emph{$r$-uniform} if it is all-odd and every positive pair $\{u,v\}$ has merge cost $\mu_W(u,v)<r$.
\end{definition}
The name refers to the following consequence, which is the form in which phase two uses the definition: two vertices joined by a positive pair have almost the same positive neighbours.
\begin{lemma}[Similar neighbourhoods]\label{lem:witness}
Let $\{u,v\}$ be a positive pair of an $r$-uniform instance. Then fewer than $r$ vertices other than $u$ and $v$ are positive to exactly one of $u$ and $v$.
\end{lemma}
\begin{proof}
Let $y\notin\{u,v\}$ be positive to exactly one of $u$ and $v$. The instance is all-odd, so $y$ is negative to the other one, and both weights have absolute value at least one. Hence the summand of $y$ in \cref{eq:merge-cost} is $\min\{|w_{uy}|,|w_{vy}|\}\ge1$. All summands are nonnegative, so the number of such vertices is at most $\mu_W(u,v)<r$.
\end{proof}

We can now state the outcome of phase one: \WCE{} on all-odd instances reduces to $r$-uniform instances, and the number of produced instances is at most $b_r^k$, with $b_r=r^{4/r}$ as in \cref{eq:base}. The threshold $r$ is part of the input, so that later sections may let it grow with the budget.
\begin{proposition}[Phase one]\label{prop:phase-one}
There is an algorithm that, given an integer $r\ge2$, an all-odd instance $W$ with $n$ vertices, and a budget $k\ge0$, runs in time $b_r^k\cdot(n+k+\tot W)^{O(1)}$ and outputs at most $b_r^k=2^{(4k\log r)/r}$ $r$-uniform instances $W_1,\dots,W_\ell$ with budgets $k_1,\dots,k_\ell$ such that
\begin{lemenum}
\item $(W,k)$ is a yes-instance if and only if $(W_i,k_i)$ is a yes-instance for some $i$, and
\item each $W_i$ has at most $n$ vertices and total weight at most $\tot W$, and $0\le k_i\le k$.
\end{lemenum}
\end{proposition}
\begin{proof}
The algorithm builds the search tree of $(W,k)$ with threshold $r$ and outputs its leaves that are $r$-uniform instances, with their budgets.

For the equivalence, consider a node that is not a leaf. By \cref{lem:safe} it is a yes-instance if and only if some produced instance is a yes-instance. Discarded instances have negative budget and are no-instances, so the node is a yes-instance if and only if one of its children is. A leaf that is not $r$-uniform is a node at which a branching rule was applied and both produced instances were discarded; by the same argument it is a no-instance. By induction on the height of the nodes, the root is a yes-instance if and only if some $r$-uniform leaf is.

The bounds on the output follow from \cref{lem:safe} and \cref{lem:tree}\ref{it:leaves}. For the running time, the tree has at most $(k+n+1)\cdot b_r^k$ nodes by \cref{lem:tree}\ref{it:leaves}. At each node, computing all merge costs, selecting a rule, and applying it takes $O(n^3)$ arithmetic operations on integers of absolute value at most $\tot W$, since total weights do not increase. The exponent $O(1)$ in the running time therefore does not depend on $r$.
\end{proof}

 \section{Phase two: solving \texorpdfstring{$r$}{r}-uniform instances}\label{sec:uniform-solve}
Phase one (\cref{prop:phase-one}) leaves us with $r$-uniform instances (\cref{def:uniform}): all weights are odd, and every positive pair has merge cost less than $r$. This section decides such instances in two steps. The first step uses no algorithm for \ThreeSAT{}. It reduces an $r$-uniform instance to a single instance in which every vertex has fewer than $4r$ positive neighbours and whose number of vertices is linear in its budget; we call such instances $r$-sparse (\cref{def:sparse,prop:to-sparse}). The second step decides an $r$-sparse instance with budget $k$ by at most $2^{k/r}$ calls of a given algorithm for \ThreeSAT{}, each on a formula with at most $c_r\cdot k$ clauses, where the number $c_r$ depends only on $r$ (\cref{prop:sparse}). A large threshold $r$ thus means few calls but long formulas; \cref{sec:proof} balances the two.

The only property of $r$-uniform instances that we use, apart from their weights being odd, is that two vertices joined by a positive pair have almost the same positive neighbours (\cref{lem:witness}). The section proceeds as follows.
\begin{compactitem}
\item \Cref{sec:dichotomy} shows that this local property has a global consequence for the graph formed by the positive pairs: in each of its connected components, either every vertex has few neighbours or every vertex has few non-neighbours. It also shows that the connected components can be solved independently.
\item \Cref{sec:dense} solves components in which every vertex has few non-neighbours. Here, splitting off a cluster is so expensive that we can afford to guess the clusters one by one; this takes time $2^{O(\sqrt k\log k)}\cdot n^{O(1)}$ whenever $r\le\sqrt k$.
\item \Cref{sec:to-sparse} combines the two previous parts into the reduction to $r$-sparse instances.
\item \Cref{sec:sparse} solves $r$-sparse instances. This is the only place where we use an algorithm for \ThreeSAT{}.
\end{compactitem}

\subsection{Sparse and dense components}\label{sec:dichotomy}
\Cref{lem:witness} compares the positive neighbours of two vertices only if the two vertices form a positive pair. We now show that this local condition restricts each connected component of the graph formed by the positive pairs as a whole (\cref{lem:dichotomy}), and that these components can be solved independently of each other (\cref{lem:components}). The first statement concerns graphs only, so we first translate the condition into graph terms.

The \emph{positive graph} $G_W$ of a weighted instance $W$ has the vertices of $W$ as its vertices and the positive pairs of $W$ as its edges; thus the positive neighbours of a vertex are its neighbours in $G_W$. For an edge $\{u,v\}$ of a graph, a \emph{witness} of $\{u,v\}$ is a vertex other than $u$ and $v$ that is adjacent to exactly one of $u$ and $v$. In these terms, \cref{lem:witness} states that every edge of the positive graph of an $r$-uniform instance has fewer than $r$ witnesses.

Having few witnesses on every edge allows two extreme kinds of connected graphs. In a path or a cycle, every edge has at most two witnesses, and every vertex has at most two neighbours. In a graph obtained from a clique by deleting a perfect matching, every edge has at most two witnesses, and every vertex has exactly one non-neighbour. The following lemma states that every connected graph with few witnesses on every edge resembles one of these two kinds: its maximum degree is bounded, or the maximum degree of its complement is.
\begin{lemma}[Degree dichotomy]\label{lem:dichotomy}
Let $r\ge1$ be an integer and let $G$ be a connected graph in which every edge has fewer than $r$ witnesses. Then $\Delta(G)<4r$ or $\Delta(\overline G)<4r$.
\end{lemma}
The proof takes a vertex $u$ of degree at least $4r$ and shows three things in turn: the closed neighbourhood $N[u]$ is almost a clique with few edges leaving it; consequently few vertices are at distance two from $u$, and each of them is adjacent to most of $N[u]$; and no vertex is at distance three. Then every vertex is adjacent to most of $N[u]$, which contains almost all vertices.
\begin{proof}
Suppose that $\Delta(G)\ge4r$, and let $u$ be a vertex of degree $h\ge4r$. Put $A:=N[u]$, so $|A|=h+1$, and let $B$ be the set of vertices outside $A$ that have a neighbour in $A$. We show that every vertex has fewer than $4r$ non-neighbours.

\emph{Step 1: vertices of $A$.} Let $v\in A\setminus\{u\}$. A vertex of $A\setminus\{u,v\}$ that is not adjacent to $v$ is adjacent to $u$ and hence is a witness of the edge $\{u,v\}$. A vertex outside $A$ that is adjacent to $v$ is not adjacent to $u$ and hence is also a witness of $\{u,v\}$. Since $\{u,v\}$ has fewer than $r$ witnesses,
\begin{align}
 \text{$v$ has fewer than $r$ non-neighbours in $A$ and fewer than $r$ neighbours outside $A$.}\label{eq:A-vertex}
\end{align}
The vertex $u$ has no neighbour outside $A$. Summing over the $h$ vertices of $A\setminus\{u\}$, fewer than $hr$ edges join $A$ to vertices outside $A$.

\emph{Step 2: vertices of $B$.} Let $w\in B$, and let $v\in A$ be a neighbour of $w$; then $v\ne u$. By \cref{eq:A-vertex}, $v$ has more than $h-r$ neighbours in $A$, because $A\setminus\{v\}$ has $h$ vertices. Each of these neighbours of $v$ differs from $w$, since $w\notin A$, so each of them that is not adjacent to $w$ is a witness of the edge $\{v,w\}$. This edge has fewer than $r$ witnesses, hence
\begin{align}
 \text{$w$ has more than $h-2r$ neighbours in $A$.}\label{eq:B-vertex}
\end{align}
Counting the edges between $A$ and $B$ from both sides, using Step~1, gives $|B|\cdot(h-2r)<hr$. Since $h\ge4r$, we have $h-2r\ge h/2>0$ and therefore
\begin{align}
 |B|<\frac{hr}{h-2r}\le2r.\label{eq:B-size}
\end{align}

\emph{Step 3: there are no further vertices.} Suppose that some vertex lies outside $A\cup B$. Since $G$ is connected, some vertex $x$ outside $A\cup B$ has a neighbour in $A\cup B$. This neighbour is not in $A$, because otherwise $x$ would belong to $B$; so it is a vertex $w\in B$, and $x$ has no neighbour in $A$. By \cref{eq:B-vertex}, $w$ has more than $h-2r\ge2r$ neighbours in $A$. Each of them is adjacent to $w$, not adjacent to $x$, and different from both, so it is a witness of the edge $\{w,x\}$. Thus $\{w,x\}$ has more than $r$ witnesses, a contradiction. Hence $V(G)=A\cup B$.

\emph{Step 4: counting non-neighbours.} The non-neighbours of $u$ are the vertices of $B$, and there are fewer than $2r$ of them by \cref{eq:B-size}. A vertex $v\in A\setminus\{u\}$ has fewer than $r$ non-neighbours in $A$ by \cref{eq:A-vertex} and at most $|B|<2r$ non-neighbours in $B$, so fewer than $3r$ in total. A vertex $w\in B$ has more than $h-2r$ neighbours among the $h+1$ vertices of $A$ by \cref{eq:B-vertex}, so at most $2r$ non-neighbours in $A$; it has at most $|B|-1\le2r-2$ non-neighbours in $B$ by \cref{eq:B-size}, so at most $4r-2$ in total. Hence $\Delta(\overline G)<4r$.
\end{proof}

We apply \cref{lem:dichotomy} to each connected component of the positive graph separately. Let $W$ be an $r$-uniform instance. A \emph{component} of $W$ is the vertex set of a connected component of its positive graph $G_W$. A component $C$ is \emph{sparse} if $\Delta(G_W[C])<4r$, and \emph{dense} otherwise. A witness of an edge is adjacent to one of its ends and thus lies in the same component, so every edge of $G_W[C]$ has fewer than $r$ witnesses in $G_W[C]$ by \cref{lem:witness}. Hence \cref{lem:dichotomy} applies to $G_W[C]$: every vertex of a sparse component has fewer than $4r$ positive neighbours, and every vertex of a dense component is positive to all but fewer than $4r$ other vertices of the component.

It remains to justify treating the components one by one. For a weighted instance $W$ with vertex set $V$ and a set $C\subseteq V$, write $W[C]$ for the submatrix of $W$ indexed by $C$. The following lemma holds for all weighted instances. Its second part is not needed for splitting into components, but it has the same proof and \cref{sec:sparse} relies on it.
\begin{lemma}[Components]\label{lem:components}
Let $W$ be a weighted instance and let $C_1,\dots,C_p$ be the vertex sets of the connected components of its positive graph $G_W$.
\begin{lemenum}
\item\label{it:comp-sum} $\OPT(W)=\sum_{i=1}^{p}\OPT(W[C_i])$.
\item\label{it:comp-connected} There is a partition $\mathcal{P}$ of cost $\OPT(W)$ such that $G_W[C]$ is connected for every cluster $C$ of $\mathcal{P}$.
\end{lemenum}
\end{lemma}
\begin{proof}
Let $\mathcal{P}$ be any partition of the vertex set of $W$, and let $\mathcal{P}^*$ be the partition obtained by replacing every cluster $C$ of $\mathcal{P}$ with the vertex sets of the connected components of $G_W[C]$. Every pair joined by $\mathcal{P}^*$ is joined by $\mathcal{P}$. A pair that is joined by $\mathcal{P}$ and separated by $\mathcal{P}^*$ lies in two different connected components of $G_W[C]$ for some cluster $C$; it is not an edge of $G_W$, so its weight is at most zero and it costs nothing under $\mathcal{P}^*$. All other pairs cost the same under both partitions. Hence $\cost_W(\mathcal{P}^*)\le\cost_W(\mathcal{P})$, and every cluster of $\mathcal{P}^*$ induces a connected subgraph of $G_W$. Taking $\mathcal{P}$ of cost $\OPT(W)$ proves \ref{it:comp-connected}.

For \ref{it:comp-sum}, take $\mathcal{P}^*$ as in \ref{it:comp-connected}. Each of its clusters induces a connected subgraph of $G_W$ and is therefore contained in one of the sets $C_i$, so for each $i$ the clusters contained in $C_i$ form a partition $\mathcal{P}_i$ of $C_i$. A pair with ends in two different sets $C_i$ and $C_j$ is separated by $\mathcal{P}^*$ and is not an edge of $G_W$, so it costs nothing. Hence $\OPT(W)=\cost_W(\mathcal{P}^*)=\sum_{i}\cost_{W[C_i]}(\mathcal{P}_i)\ge\sum_i\OPT(W[C_i])$. Conversely, the union of partitions of cost $\OPT(W[C_i])$ of the sets $C_i$ is a partition of the vertex set of $W$. It separates every pair with ends in two different sets, and these pairs cost nothing as before, so its cost is $\sum_i\OPT(W[C_i])$.
\end{proof}

\subsection{Dense components}\label{sec:dense}
By \cref{sec:dichotomy}, every vertex of a dense component is positive to all but fewer than $4r$ other vertices of the component. This section decides instances with this property in subexponential time, without using an algorithm for \ThreeSAT{}. The idea is that such an instance leaves little choice: a partition with at least two clusters has a cluster $C$ with at most half of the vertices, and every vertex of $C$ is then separated from almost half of all vertices, to nearly all of which it is positive. So either the whole vertex set is one cluster, or the partition pays about $|C|\cdot n/2$ for a cluster $C$ that is one of at most $n^{|C|}$ sets. The budget thus allows for few vertices in such clusters, and we can afford to guess them.

The argument needs neither odd weights nor connectivity, so we state it for every weighted instance $W$ with $\Delta(\overline{G_W})<4r$; that is, every vertex has a weight of at most zero to fewer than $4r$ other vertices. This property is inherited by $W[C']$ for every vertex set $C'$, which allows us to recurse. We need two more pieces of notation. For a weighted instance $W$ with vertex set $V$ and a set $C\subseteq V$, put $W-C:=W[V\setminus C]$, and define the \emph{splitting cost} of $C$ as
\begin{align}
 \sigma_W(C):=\sum_{\{u,v\}\subseteq C}\pos{-w_{uv}}+\sum_{u\in C,\ y\in V\setminus C}\pos{w_{uy}}.\label{eq:split-cost}
\end{align}
This is what a partition having $C$ as a cluster pays for the pairs that meet $C$: the pairs inside $C$ are joined and the pairs between $C$ and $V\setminus C$ are separated. All other pairs avoid $C$. Hence, for every partition $\mathcal{P}$ of $V$ that has $C$ as a cluster,
\begin{align}
 \cost_W(\mathcal{P})=\sigma_W(C)+\cost_{W-C}(\mathcal{P}\setminus\{C\}).\label{eq:split-off}
\end{align}
For example, $\sigma_W(V)$ is the cost of the partition $\{V\}$, and $\sigma_W(C)\ge0$ for every $C$.

The next lemma makes the price of a cluster with at most half of the vertices precise.
\begin{lemma}[Splitting is expensive]\label{lem:split-expensive}
Let $r\ge1$ be an integer, let $W$ be a weighted instance with $n$ vertices and $\Delta(\overline{G_W})<4r$, and let $C$ be a set of at most $n/2$ vertices. Then $\sigma_W(C)\ge|C|\cdot(n/2-4r)$.
\end{lemma}
\begin{proof}
Let $u\in C$. At least $n/2$ vertices lie outside $C$, and fewer than $4r$ of them are non-neighbours of $u$ in $G_W$. So $u$ is positive to more than $n/2-4r$ vertices outside $C$, and each of these pairs has an integer weight of at least one. The second sum of \cref{eq:split-cost} is therefore at least $|C|\cdot(n/2-4r)$, and the first sum is nonnegative.
\end{proof}

The algorithm $\Dense(W,k)$ takes a weighted instance $W$ with vertex set $V$, $n:=|V|$, and $\Delta(\overline{G_W})<4r$, and an integer $k$. It proceeds as follows.
\begin{compactenum}[1.]
\item If $k<0$, answer no.
\item If $\sigma_W(V)\le k$, answer yes.
\item For every nonempty set $C\subseteq V$ with $|C|\le n/2$ and $|C|\cdot(n/2-4r)\le k$: if $\Dense(W-C,k-\sigma_W(C))$ answers yes, answer yes.
\item Answer no.
\end{compactenum}
Whether a set $C$ is tried in Step~3 depends only on its size, so the algorithm lists the sets to try without inspecting any others.

\begin{proposition}[Dense instances]\label{prop:dense}
Let $r\ge1$ be an integer. Given a weighted instance $W$ with $n$ vertices and $\Delta(\overline{G_W})<4r$ and an integer $k\ge0$, the algorithm \Dense\ decides whether $\OPT(W)\le k$ in time
\[
 (4k+16r)^{4\sqrt k+16r+1}\cdot(n+\tot W)^{O(1)}.
\]
If $2\le r\le\sqrt k$, this is $2^{O(\sqrt k\log k)}\cdot(n+\tot W)^{O(1)}$, where the constants hidden in the $O$-notation do not depend on $r$.
\end{proposition}
The proof of the running time follows one path of the recursion and shows that all sets guessed along it together have $O(\sqrt k+r)$ vertices. As long as many vertices remain, \cref{lem:split-expensive} makes every guessed vertex expensive; once few vertices remain, few can be guessed.
\begin{proof}
\emph{Correctness.} We use induction on $n$. If $k<0$, then no partition has cost at most $k$, and Step~1 is correct. Let $k\ge0$. If the algorithm answers yes in Step~2, then the partition $\{V\}$ has cost at most $k$ (the empty partition if $V$ is empty). If it answers yes in Step~3 for a set $C$, then by induction $W-C$ has a partition of cost at most $k-\sigma_W(C)$; adding the cluster $C$ gives a partition of $W$ of cost at most $k$ by \cref{eq:split-off}. Conversely, let $\mathcal{P}$ be a partition with $\cost_W(\mathcal{P})\le k$. If $\mathcal{P}$ has at most one cluster, then $\cost_W(\mathcal{P})=\sigma_W(V)$ and Step~2 answers yes. Otherwise let $C$ be a cluster of $\mathcal{P}$ with the fewest vertices; then $1\le|C|\le n/2$. By \cref{eq:split-off} we have $\sigma_W(C)\le\cost_W(\mathcal{P})\le k$, so $|C|\cdot(n/2-4r)\le k$ by \cref{lem:split-expensive}, and Step~3 tries $C$. Again by \cref{eq:split-off}, the partition $\mathcal{P}\setminus\{C\}$ of $W-C$ has cost at most $k-\sigma_W(C)$, so the recursive call answers yes by induction.

\emph{The recursion tree.} Consider the tree of recursive calls; its root is the call $\Dense(W,k)$. If $n/2-4r>k$, then Step~3 tries no set and the tree has one node. So assume
\begin{align}
 n\le2k+8r.\label{eq:dense-n}
\end{align}
A node of the tree is determined by the sequence $(C_1,\dots,C_m)$, $m\ge0$, of the sets tried on the path from the root to it. These sets are nonempty and pairwise disjoint. We claim that
\begin{align}
 |C_1|+\dots+|C_m|\le4\sqrt k+16r.\label{eq:dense-total}
\end{align}
For $1\le i\le m$, let $n_i$ and $k_i$ be the number of vertices and the budget of the call that tries $C_i$. Then $n_1>n_2>\dots>n_m$, and $k_i=k-\sum_{j<i}\sigma_j$, where $\sigma_j$ is the splitting cost of $C_j$ in the instance of the call that tries it. Put $q:=\max\{16r,2\sqrt k\}$. If $n_i\ge q$, then $n_i/2-4r\ge n_i/4$ since $n_i\ge16r$.

Let $i^*$ be the largest index with $n_{i^*}\ge q$, if there is one. For $j<i^*$ we have $n_j\ge q$ and thus $\sigma_j\ge|C_j|\cdot n_j/4\ge|C_j|\cdot q/4$ by \cref{lem:split-expensive}. Since Step~3 tried $C_{i^*}$,
\[
 |C_{i^*}|\cdot\frac q4\le|C_{i^*}|\cdot\Bigl(\frac{n_{i^*}}2-4r\Bigr)\le k_{i^*}=k-\sum_{j<i^*}\sigma_j\le k-\sum_{j<i^*}|C_j|\cdot\frac q4.
\]
Hence $\sum_{j\le i^*}|C_j|\le4k/q\le2\sqrt k$; the last inequality holds because $q\ge2\sqrt k$. The sets $C_j$ with $j>i^*$, or all sets if there is no index $i^*$, are disjoint subsets of a set of fewer than $q$ vertices, so they have fewer than $q\le16r+2\sqrt k$ vertices in total. This proves \cref{eq:dense-total}.

It remains to count sequences. A sequence of pairwise disjoint nonempty subsets of $V$ with $j$ vertices in total is determined by a list of $j$ distinct vertices, consisting of the vertices of the first set in some order, then those of the second set, and so on, together with the positions in the list at which a new set starts. There are at most $n^j$ such lists and at most $2^j$ choices of positions. By \cref{eq:dense-total}, and since $n\ge1$ may be assumed, the number of nodes is at most
\[
 \sum_{j=0}^{\lfloor4\sqrt k+16r\rfloor}(2n)^j\le(2n)^{4\sqrt k+16r+1}\le(4k+16r)^{4\sqrt k+16r+1},
\]
where the last inequality is \cref{eq:dense-n}.

\emph{Running time.} Every set tried in Step~3 gives a child in the recursion tree, so the work at a node, apart from the recursive calls, is polynomial in $n$ and $\tot W$ per child. Hence the total running time is the number of nodes times a polynomial in $n+\tot W$. If $2\le r\le\sqrt k$, then $k\ge4$, $16r\le16k$, and $4\sqrt k+16r+1\le21\sqrt k$, so the number of nodes is at most $(20k)^{21\sqrt k}=2^{O(\sqrt k\log k)}$.
\end{proof}

\subsection{From \texorpdfstring{$r$}{r}-uniform instances to sparse instances}\label{sec:to-sparse}
By \cref{sec:dichotomy}, an $r$-uniform instance (\cref{def:uniform}) splits into dense and sparse components that can be solved independently, and the algorithm \Dense{} of \cref{prop:dense} computes the optimum of a dense component. What remains is the union of the sparse components, an all-odd instance in which every vertex has fewer than $4r$ positive neighbours. Such an instance can still have any number of vertices, whereas the encoding into formulas in \cref{sec:sparse} needs a number of vertices that is linear in the budget. We show that the remaining instance has this property unless its optimum exceeds the budget, once the components whose positive graph is complete are removed (\cref{lem:sparse-size}); this bound takes the role that a problem kernel would otherwise play. The outcome is a reduction from $r$-uniform instances to instances of the following kind (\cref{prop:to-sparse}).
\begin{definition}\label{def:sparse}
Let $r\ge1$ be an integer. An \emph{$r$-sparse instance} is a pair $(W,k)$ of an all-odd instance $W$ and an integer $k\ge0$ such that
\begin{lemenum}
\item\label{it:sparse-degree} $\Delta(G_W)<4r$, that is, every vertex has fewer than $4r$ positive neighbours,
\item\label{it:sparse-noclique} no connected component of the positive graph $G_W$ is a complete graph, and
\item\label{it:sparse-size} $W$ has at most $16rk$ vertices.
\end{lemenum}
\end{definition}
For example, let $G$ be a path with three vertices and recall the instance $W_G$ from \cref{sec:model}. Then $(W_G,1)$ is a $1$-sparse instance: $W_G$ is all-odd, its positive graph is $G$, which is connected, not complete, and has maximum degree $2<4$, and $3\le16$. The pair $(W_G,0)$ is not $1$-sparse because of \ref{it:sparse-size}; in general, an $r$-sparse instance with budget zero has no vertices. Throughout \cref{sec:to-sparse,sec:sparse}, $r\ge1$ is an integer.

\paragraph{Negative pairs in large sets.}
Both the bound on the number of vertices (\cref{lem:sparse-size}) and, in \cref{sec:sparse}, the bound on the size of the clusters that a formula chooses rest on the following count: a set with many vertices contains many negative pairs, because every vertex has few positive neighbours.
\begin{lemma}[Negative pairs in large sets]\label{lem:neg-inside}
Let $W$ be an all-odd instance with $\Delta(G_W)<4r$, and let $C$ be a set of more than $8r$ vertices. Then
\[
 \sum_{\{u,v\}\subseteq C}\pos{-w_{uv}}>\frac{|C|^2}{4}.
\]
\end{lemma}
\begin{proof}
Every vertex of $C$ is positive to fewer than $4r$ vertices of $C$ and therefore negative to at least $|C|-4r>|C|/2$ of them, since all weights are nonzero. Hence $C$ contains more than $|C|^2/4$ negative pairs, and each of them has weight at most $-1$.
\end{proof}

\paragraph{Few vertices.}
The next lemma bounds the number of vertices of an instance whose positive graph is connected and has small maximum degree by its optimum. A partition into few clusters has a cluster with many vertices, which is expensive by \cref{lem:neg-inside}; a partition into many clusters separates many positive pairs, because the positive graph is connected.
\begin{lemma}[Sparse instances are small]\label{lem:sparse-size}
Let $W$ be an all-odd instance with $n$ vertices such that $G_W$ is connected and $\Delta(G_W)<4r$. Then $\OPT(W)\ge n/(8r)-1$. If, in addition, $G_W$ is not a complete graph, then $\OPT(W)\ge n/(16r)$.
\end{lemma}
\begin{proof}
Let $\mathcal{P}$ be a partition with $p$ clusters, and let $n'$ be the number of vertices in clusters with at most $8r$ vertices. There are at least $n'/(8r)$ such clusters, so $p\ge n'/(8r)$. Consider the graph whose vertices are the clusters of $\mathcal{P}$ and in which two clusters are adjacent if some positive pair has one vertex in each. This graph is connected because $G_W$ is, so it has at least $p-1$ edges, and each edge accounts for a different positive pair separated by $\mathcal{P}$. Each such pair costs at least one. A cluster $C$ with more than $8r$ vertices costs more than $|C|^2/4\ge|C|/(8r)$ for the pairs inside it by \cref{lem:neg-inside}. Hence
\[
 \cost_W(\mathcal{P})\ge(p-1)+\frac{n-n'}{8r}\ge\frac{n'}{8r}-1+\frac{n-n'}{8r}=\frac n{8r}-1.
\]
For the second statement, suppose that some partition has cost zero. It separates no positive pair, so it has a single cluster because $G_W$ is connected; and it joins no negative pair, so every pair is positive, as no weight is zero. Thus $G_W$ is complete. Hence $\OPT(W)\ge1$ if $G_W$ is not complete. If $n\le16r$, this gives $\OPT(W)\ge n/(16r)$; if $n>16r$, then $n/(8r)-1\ge n/(16r)$.
\end{proof}

\paragraph{The reduction.}
We can now state the reduction. It is the part of phase two that needs no algorithm for \ThreeSAT{}.
\begin{proposition}[Reduction to sparse instances]\label{prop:to-sparse}
There is an algorithm that, given an integer $r\ge1$, an $r$-uniform instance $W$ with $n$ vertices, and an integer $k\ge0$, runs in time
\[
 (4k+16r)^{4\sqrt k+16r+1}\cdot(n+\tot W)^{O(1)}
\]
and either reports correctly that $\OPT(W)>k$ or outputs an $r$-sparse instance $(W',k')$ such that
\begin{lemenum}
\item\label{it:to-sparse-equiv} $\OPT(W)\le k$ if and only if $\OPT(W')\le k'$, and
\item\label{it:to-sparse-size} $k'\le k$, and $W'$ has at most $n$ vertices and total weight at most $\tot W$.
\end{lemenum}
\end{proposition}
The algorithm computes the optimum of every dense component, subtracts these optima from the budget, and outputs what is left of the sparse components.
\begin{proof}
Compute the components of $W$ (\cref{sec:dichotomy}). Let $C_1,\dots,C_p$ be the dense components and let $S$ be the union of the sparse components. The connected components of $G_W[S]$ are the sparse components, so applying \cref{lem:components}\ref{it:comp-sum} to $W$ and to $W[S]$ gives
\begin{align}
 \OPT(W)=\OPT(W[S])+\sum_{i=1}^p\OPT(W[C_i]).\label{eq:combine}
\end{align}

\emph{Dense components.} By \cref{sec:dichotomy}, every instance $W[C_i]$ satisfies $\Delta(\overline{G_{W[C_i]}})<4r$. For each $i$, call the algorithm \Dense{} of \cref{prop:dense} on $W[C_i]$ with the budgets $0,1,\dots,k$ in turn until it answers yes; this determines $\OPT(W[C_i])$ or shows that it exceeds $k$. If some dense component has optimum more than $k$, or if the sum $k''$ of the optima of the dense components exceeds $k$, report that $\OPT(W)>k$, which is correct by \cref{eq:combine}. Otherwise put $k':=k-k''$; then $0\le k'\le k$, and by \cref{eq:combine}, $\OPT(W)\le k$ if and only if $\OPT(W[S])\le k'$.

\emph{Sparse components.} Let $S'$ be the union of those sparse components $C$ for which $G_W[C]$ is not a complete graph, and put $W':=W[S']$. A sparse component $C$ for which $G_W[C]$ is complete has only positive pairs, so the partition $\{C\}$ of $C$ costs nothing and $\OPT(W[C])=0$. Hence \cref{lem:components}\ref{it:comp-sum}, applied to $W[S]$ and to $W'$, gives $\OPT(W[S])=\OPT(W')$. Every connected component of $G_{W'}$ is a sparse component $C$ of $W$ with $G_W[C]$ not complete, and $W[C]$ is all-odd with $\Delta(G_{W[C]})<4r$, so $\OPT(W[C])\ge|C|/(16r)$ by \cref{lem:sparse-size}. Summing over these components gives $\OPT(W')\ge|S'|/(16r)$, again by \cref{lem:components}\ref{it:comp-sum}. If $|S'|>16rk'$, report that $\OPT(W)>k$; this is correct because then $\OPT(W[S])=\OPT(W')>k'$. Otherwise output $(W',k')$.

\emph{The output.} The instance $W'$ is all-odd because $W$ is. Its positive graph is $G_W[S']$, whose connected components are sparse components of $W$ that are not complete; hence \ref{it:sparse-degree} and \ref{it:sparse-noclique} of \cref{def:sparse} hold, and \ref{it:sparse-size} holds because $|S'|\le16rk'$. So $(W',k')$ is $r$-sparse. Statement \ref{it:to-sparse-equiv} was shown above, and \ref{it:to-sparse-size} holds because $W'$ is a submatrix of $W$.

\emph{Running time.} There are $p\le n$ dense components and at most $k+1$ calls of \Dense{} for each, with budgets at most $k$, and each call takes time $(4k+16r)^{4\sqrt k+16r+1}\cdot(n+\tot W)^{O(1)}$ by \cref{prop:dense}. Everything else takes time polynomial in $n+\tot W$, and $k+1\le4k+16r$.
\end{proof}

\subsection{Deciding sparse instances}\label{sec:sparse}
It remains to decide $r$-sparse instances $(W,k)$ (\cref{def:sparse}): every vertex has fewer than $4r$ positive neighbours, no connected component of the positive graph is complete, and there are at most $16rk$ vertices. Unlike in dense instances (\cref{sec:dense}), a single cluster can now be split off cheaply, so guessing clusters one by one is too slow. Instead we encode the choice of the whole partition into \ThreeSAT{} formulas and decide them by a given algorithm for \ThreeSAT{}. The running time of that algorithm is measured in the number of clauses (\cref{sec:model}), so the formulas must have a number of clauses that is linear in $k$. The factor of linearity depends on $r$, and we keep track of it, because \cref{sec:proof} lets $r$ grow with $k$. Two observations make such formulas possible; both refer to the abbreviation $\lambda:=256r^2$.
\begin{compactitem}
\item Some optimal partition consists of clusters that are connected in the positive graph (\cref{lem:components}). Since every vertex has fewer than $4r$ positive neighbours, every vertex lies in at most $(4r)^{2\lambda}$ connected sets with at most $\lambda$ vertices. We call these sets \emph{candidates}, up to a further restriction from the next item. One variable for each candidate, together with the constraint that every vertex lies in exactly one chosen candidate, gives a number of variables and clauses that is linear in the number of vertices and hence in $k$. Clusters with more than $\lambda$ vertices are expensive, so a partition within budget has few vertices in such clusters, and we guess them (\cref{lem:candidates}).
\item The formula must also state that the chosen partition costs at most $k$. The cost of a partition is a sum with one term for each cluster, which we call the \emph{share} of the cluster. If every share is bounded in terms of $r$, the formula can count the shares in unary: a variable whose set has share $a$ is counted $a$ times, and counting a number of bits that is linear in $k$ takes a circuit with linearly many gates (\cref{lem:threshold}). Without the bound, the sum would involve numbers with about $\log k$ binary digits each, and the circuit would be too large. A cluster whose share exceeds the bound is expensive, so a partition within budget has few such clusters; we guess them along with the previous ones.
\end{compactitem}
The guessed clusters have at most $8k/\lambda$ vertices in total, and this leaves at most $2^{k/r}$ guesses.

\paragraph{Shares and candidates.}
We now prepare the encoding of partitions. To express the cost of a partition as a sum over its clusters, define the \emph{share} of a set $C\subseteq V$ in a weighted instance $W$ with vertex set $V$ as
\begin{align}
 \alpha_W(C):=2\cdot\sum_{\{u,v\}\subseteq C}\pos{-w_{uv}}+\sum_{u\in C,\ y\in V\setminus C}\pos{w_{uy}}.\label{eq:share}
\end{align}
It differs from the splitting cost \cref{eq:split-cost} in the factor $2$, which makes the shares of the clusters of a partition add up to twice its cost.
\begin{lemma}[Shares]\label{lem:shares}
Let $W$ be a weighted instance with vertex set $V$. For every partition $\mathcal{P}$ of $V$,
\[
 2\cdot\cost_W(\mathcal{P})=\sum_{C\in\mathcal{P}}\alpha_W(C).
\]
\end{lemma}
\begin{proof}
A pair joined by $\mathcal{P}$ lies inside exactly one cluster, and the share of this cluster counts the cost of the pair twice. A pair separated by $\mathcal{P}$ has its two vertices in two different clusters, whose shares count the cost of the pair once each.
\end{proof}

Throughout the rest of this section, put $\lambda:=256r^2$. The arguments below use only that $\lambda$ is an integer with $\lambda\ge8r$, except for the count of guesses in the proof of \cref{prop:sparse}, which needs $\lambda$ to be this large. A \emph{candidate} of an all-odd instance $W$ with $\Delta(G_W)<4r$ is a nonempty set $C$ of vertices such that $G_W[C]$ is connected and $\alpha_W(C)\le\lambda^2/2$. Candidates are the clusters that the formula will choose by its variables; the following lemma states that they are small and few around each vertex, and that the remaining clusters of a partition within budget have few vertices.
\begin{lemma}[Candidates]\label{lem:candidates}
Let $W$ be an all-odd instance with $\Delta(G_W)<4r$, and recall that $\lambda=256r^2$.
\begin{lemenum}
\item\label{it:cand-size} Every candidate has at most $\lambda$ vertices.
\item\label{it:cand-number} Every vertex lies in at most $(4r)^{2\lambda}$ candidates.
\item\label{it:cand-rest} Let $k\ge0$, and let $\mathcal{P}$ be a partition with $\cost_W(\mathcal{P})\le k$ all of whose clusters induce connected subgraphs of $G_W$. Then the clusters of $\mathcal{P}$ that are not candidates have at most $8k/\lambda$ vertices in total.
\end{lemenum}
\end{lemma}
\begin{proof}
\ref{it:cand-size} A set $C$ with more than $\lambda\ge8r$ vertices has share $\alpha_W(C)>|C|^2/2>\lambda^2/2$ by \cref{lem:neg-inside} and the definition \cref{eq:share}.

\ref{it:cand-number} We describe every candidate that contains a given vertex $v$ by a short sequence of symbols. Fix, for every vertex, an order of its positive neighbours; there are at most $4r-1$ of them. Let $C$ be a candidate with $v\in C$, and put $s:=|C|$, so $s\le\lambda$ by \ref{it:cand-size}. The graph $G_W[C]$ is connected and therefore has a spanning tree, which we root at $v$. A depth-first traversal of this tree is a walk that starts at $v$, visits every vertex of $C$, and passes along each of the $s-1$ edges of the tree exactly twice, once away from the root and once back; so it has $2(s-1)$ steps. We record a step from a vertex $x$ to a child of $x$ by the position of this child in the order of the positive neighbours of $x$, a number between $1$ and $4r-1$, and a step from $x$ back to its parent by the symbol $0$. The resulting sequence of $2(s-1)$ symbols determines the walk step by step, because the parent of a vertex is the vertex from which the walk first reached it. Hence the sequence determines $C$, the set of vertices that the walk visits. So the number of candidates that contain $v$ is at most the number of sequences of at most $2\lambda-2$ symbols from $\{0,\dots,4r-1\}$, which is $\sum_{j=0}^{2\lambda-2}(4r)^j\le(4r)^{2\lambda-1}$.

\ref{it:cand-rest} Every cluster of $\mathcal{P}$ induces a connected subgraph of $G_W$ by assumption. So a cluster of $\mathcal{P}$ that is not a candidate has share more than $\lambda^2/2$, by the definition of candidates. By \cref{lem:shares}, the shares of the clusters of $\mathcal{P}$ sum to at most $2k$, so at most $4k/\lambda^2$ clusters of $\mathcal{P}$ are not candidates. Those of them that have at most $\lambda$ vertices have at most $4k/\lambda$ vertices in total. A cluster $C$ with more than $\lambda$ vertices has share more than $|C|^2/2>|C|\cdot\lambda/2$ by \cref{lem:neg-inside}, so the clusters with more than $\lambda$ vertices have fewer than $4k/\lambda$ vertices in total.
\end{proof}

\paragraph{Formulas.}
Recall 3-CNF formulas from \cref{sec:model}. We describe formulas by \emph{constraints}: a constraint on a set $X$ of variables is a set of assignments of $X$, those that it allows. Two textbook transformations turn constraints into 3-CNF formulas. First, a constraint on $g$ variables is equivalent to the conjunction of at most $2^g$ clauses with $g$ literals each, one clause for each assignment that is not allowed, which is the only assignment falsifying that clause. Second, a clause $\ell_1\vee\dots\vee\ell_g$ with $g>3$ literals can be replaced by the clauses $\ell_1\vee\ell_2\vee a_1$, $\neg a_{i-1}\vee\ell_{i+1}\vee a_i$ for $2\le i\le g-3$, and $\neg a_{g-3}\vee\ell_{g-1}\vee\ell_g$, where $a_1,\dots,a_{g-3}$ are new variables: an assignment of the old variables satisfies the long clause if and only if it can be extended to the new variables so that all the new clauses are satisfied. Applying the first transformation and then the second to each of its clauses gives the following.
\begin{lemma}[Constraints as formulas]\label{lem:constraint}
A constraint on $g$ variables can be replaced by a 3-CNF formula on these variables and at most $g\cdot2^g$ new variables, with at most $g\cdot2^g$ clauses, such that an assignment of the $g$ variables is allowed by the constraint if and only if it extends to a satisfying assignment of the formula.
\end{lemma}

The number of clauses in \cref{lem:constraint} is exponential in $g$. This is harmless for constraints on a bounded number of variables, such as the adders in \cref{lem:threshold} below. The constraint that exactly one candidate containing a given vertex is chosen, however, concerns up to $(4r)^{2\lambda}$ variables (\cref{lem:candidates}\ref{it:cand-number}), and we want the size of our formulas to be polynomial in this number. For such constraints we use the following encoding of linear size, which scans the variables in a fixed order and remembers whether a true one has been seen.
\begin{lemma}[Exactly one]\label{lem:exactly-one}
Let $x_1,\dots,x_g$ be $g\ge0$ distinct Boolean variables. There is a 3-CNF formula on these variables and $g+1$ new variables, with $4g+2$ clauses, such that an assignment of $x_1,\dots,x_g$ extends to a satisfying assignment of the formula if and only if it makes exactly one of $x_1,\dots,x_g$ true.
\end{lemma}
\begin{proof}
Let $p_0,\dots,p_g$ be the new variables; the intended meaning of $p_i$ is that at least one of $x_1,\dots,x_i$ is true. The formula consists of the clause $\neg p_0$, the clause $p_g$, and, for every $i$ with $1\le i\le g$, the four clauses
\[
 \neg p_{i-1}\vee p_i,\qquad\neg x_i\vee p_i,\qquad\neg p_i\vee p_{i-1}\vee x_i,\qquad\neg p_{i-1}\vee\neg x_i.
\]
Fix an assignment of $x_1,\dots,x_g$. The clause $\neg p_0$ and the first three clauses for every $i$ are satisfied if and only if $p_0$ is false and $p_i$ has the value of $p_{i-1}\vee x_i$ for every $i$; by induction on $i$, this is the case if and only if every $p_i$ has its intended meaning. So exactly one extension of the assignment satisfies these clauses. Under this extension, the fourth clause for $i$ is violated if and only if $x_i$ is true and one of $x_1,\dots,x_{i-1}$ is true. Hence all fourth clauses are satisfied if and only if at most one of $x_1,\dots,x_g$ is true, and the clause $p_g$ is satisfied if and only if at least one of them is true. For $g=0$ the formula consists of the clauses $\neg p_0$ and $p_0$ and is unsatisfiable, as required.
\end{proof}

\paragraph{Thresholds.}
The cost of the partition chosen by the formula will be a sum of shares, and the formula has to state that this sum does not exceed a threshold. The next lemma provides this for sums of bits, with a number of variables that is linear in the number of bits. We allow a variable to occur several times in the sum; this is how we will give a variable an integer coefficient.
\begin{lemma}[Thresholds]\label{lem:threshold}
There is an integer $c_0\ge1$ such that the following holds. Let $y_1,\dots,y_m$ be a list of $m\ge1$ Boolean variables, not necessarily distinct, and let $t\ge0$ be an integer. One can construct in time polynomial in $m$ a 3-CNF formula on these variables and at most $c_0\cdot m$ further variables, with at most $c_0\cdot m$ clauses, such that an assignment $\beta$ of $y_1,\dots,y_m$ extends to a satisfying assignment of the formula if and only if $\sum_{i=1}^m\beta(y_i)\le t$.
\end{lemma}
The proof builds the usual circuit that counts in binary: three bits of the same significance are replaced by a sum bit of that significance and a carry bit of the next one.
\begin{proof}
We maintain a list of \emph{bits}, each of which is a variable with a \emph{level}, a nonnegative integer. Under an assignment, the \emph{value} of the list is the sum, over its bits, of $2^i$ times the value of the variable, where $i$ is the level of the bit. Initially the list consists of $y_1,\dots,y_m$, all at level $0$. A \emph{full adder} removes three bits $a,a',a''$ of the same level $i$ from the list, introduces two new variables $s$ and $c$ with the constraint $a+a'+a''=s+2c$, and adds $s$ at level $i$ and $c$ at level $i+1$ to the list. A \emph{half adder} does the same with two bits $a,a'$ and the constraint $a+a'=s+2c$. In both cases the values of the removed bits determine the values of $s$ and $c$, and the value of the list does not change.

We process the levels $i=0,1,2,\dots$ in this order as long as bits of level at least $i$ exist. At level $i$, we apply full adders as long as there are at least three bits of level $i$, and then one half adder if there are exactly two. Afterwards, at most one bit of level $i$ is left, and no bit of level $i$ is created later. Let $N_i$ be the number of bits of level $i$ at the moment when the processing of level $i$ starts. A full adder lowers the number of bits of level $i$ by two and a half adder lowers it by one, so at most $N_i/2$ adders are applied at level $i$. Each of them creates one bit of level $i+1$, so $N_{i+1}\le N_i/2$. With $N_0=m$ we obtain $N_i\le m/2^i$. Hence at most $\sum_{i\ge0}m/2^{i+1}\le m$ adders are applied in total, and in the end the list has at most one bit at each level $i\le\lfloor\log m\rfloor$ and no other bits. Put $\ell:=\lfloor\log m\rfloor$. For each level $i\le\ell$ without a bit we add a new variable as a bit of level $i$, with a clause stating that it is false. Let $z_0,\dots,z_\ell$ be the final bits. Every assignment $\beta$ of $y_1,\dots,y_m$ extends in exactly one way to an assignment that satisfies all constraints so far, and under this assignment
\begin{align}
 \sum_{i=0}^{\ell}2^i\cdot z_i=\sum_{i=1}^m\beta(y_i).\label{eq:binary}
\end{align}

It remains to compare the left-hand side of \cref{eq:binary} with $t$. If $t\ge2^{\ell+1}$, the comparison holds for all values and we add nothing. Otherwise write $t=\sum_{i=0}^\ell2^i\cdot t_i$ with $t_i\in\{0,1\}$, and for every $i$ with $t_i=0$ add the clause
\begin{align}
 \neg z_i\vee\bigvee_{j>i,\ t_j=1}\neg z_j.\label{eq:compare}
\end{align}
We claim that these clauses hold if and only if $\sum_i2^iz_i\le t$. If $\sum_i2^iz_i>t$, let $i$ be the largest index with $z_i\ne t_i$. Then $z_i=1$ and $t_i=0$, and $z_j=t_j$ for all $j>i$, so the clause \cref{eq:compare} for $i$ is violated. Conversely, if the clause for some $i$ is violated, then $z_i=1$, $t_i=0$, and $z_j\ge t_j$ for all $j>i$. Then $\sum_{j\ge i}2^jz_j\ge2^i+\sum_{j>i}2^jt_j>\sum_{j<i}2^jt_j+\sum_{j>i}2^jt_j=t$, where the strict inequality uses $\sum_{j<i}2^j<2^i$.

Finally we turn everything into a 3-CNF formula. Each adder constraint concerns at most five variables and becomes a 3-CNF formula with a constant number of clauses and new variables by \cref{lem:constraint}. There are at most $\ell+1$ clauses \cref{eq:compare}, each with at most $\ell+1$ literals; the second transformation above replaces them by at most $(\ell+1)^2$ clauses with at most $(\ell+1)^2$ new variables, and $(\ell+1)^2\le(1+\log m)^2=O(m)$. In total the formula has $O(m)$ variables besides $y_1,\dots,y_m$ and $O(m)$ clauses, with constants that do not depend on anything; we let $c_0$ be an integer that is at least both constants. By \cref{eq:binary}, the claim on the clauses \cref{eq:compare}, and the properties of the two transformations, an assignment $\beta$ extends to a satisfying assignment if and only if $\sum_{i=1}^m\beta(y_i)\le t$.
\end{proof}

\paragraph{The formula of a guess.}
We now combine candidates and thresholds. Let $W$ be an all-odd instance with vertex set $V$ and $\Delta(G_W)<4r$, and let $k\ge0$ be an integer. A \emph{guess} is a pair $(U,\mathcal{Q})$ of a set $U\subseteq V$ and a partition $\mathcal{Q}$ of $U$; it stands for the clusters that are not chosen by the formula. The formula $\varphi_{U,\mathcal{Q}}$ of the guess has one variable $x_C$ for each candidate $C$ of $W$ with $C\subseteq V\setminus U$, and it consists of the following parts, turned into 3-CNF formulas with further variables as stated.
\begin{compactitem}
\item \emph{Partition.} For every vertex $v\in V\setminus U$, the formula of \cref{lem:exactly-one} for the variables $x_C$ with $v\in C$, with new variables of its own; it states that exactly one of these variables is true.
\item \emph{Budget.} Put $t:=2k-\sum_{C\in\mathcal{Q}}\alpha_W(C)$. If $t<0$, the empty clause, which makes the formula unsatisfiable. Otherwise the formula of \cref{lem:threshold} for the threshold $t$ and the list that contains each variable $x_C$ exactly $\alpha_W(C)$ times; if this list is empty, nothing.
\end{compactitem}
Shares are computed in $W$, not in $W-U$, so they account for the pairs between $U$ and $V\setminus U$. The size of the formula is governed by the number
\begin{align}
 c_r:=16\cdot c_0\cdot r\cdot\lambda^2\cdot(4r)^{2\lambda},\qquad\text{where }\lambda=256r^2\label{eq:cr}
\end{align}
and $c_0$ is the integer of \cref{lem:threshold}. Taking logarithms gives $\log c_r=\log(16c_0)+\log r+2\log\lambda+2\lambda\log(4r)$, so
\begin{align}
 c_r=2^{O(r^2\log r)}\quad\text{for }r\ge2.\label{eq:cr-growth}
\end{align}
\begin{lemma}[Formula of a guess]\label{lem:formula}
Let $W$ and $k$ be as above, and let $(U,\mathcal{Q})$ be a guess.
\begin{lemenum}
\item\label{it:formula-sat} The formula $\varphi_{U,\mathcal{Q}}$ is satisfiable if and only if there is a partition $\mathcal{P}$ of $V$ with $\cost_W(\mathcal{P})\le k$ such that $\mathcal{Q}\subseteq\mathcal{P}$ and all clusters in $\mathcal{P}\setminus\mathcal{Q}$ are candidates.
\item\label{it:formula-size} The formula $\varphi_{U,\mathcal{Q}}$ has at most $c_r\cdot\max\{n,1\}/(16r)$ clauses, where $n:=|V|$, and it can be constructed in time $(c_r\cdot(n+\tot W+1))^{O(1)}$, where the exponent $O(1)$ does not depend on $r$.
\end{lemenum}
\end{lemma}
\begin{proof}
\ref{it:formula-sat} By \cref{lem:exactly-one} and \cref{lem:threshold}, the formula is satisfiable if and only if some assignment of the variables $x_C$ satisfies the partition constraints and makes the sum of the shares $\alpha_W(C)$ over the candidates $C$ with $x_C$ true at most $t$, where $t\ge0$. The partition constraints state that the candidates $C$ with $x_C$ true form a partition $\mathcal{S}$ of $V\setminus U$. Then $\mathcal{P}:=\mathcal{Q}\cup\mathcal{S}$ is a partition of $V$, and by \cref{lem:shares}
\[
 2\cdot\cost_W(\mathcal{P})=\sum_{C\in\mathcal{Q}}\alpha_W(C)+\sum_{C\in\mathcal{S}}\alpha_W(C),
\]
which is at most $2k$ if and only if the second sum is at most $t$. Conversely, a partition $\mathcal{P}$ as in the statement gives the assignment that makes $x_C$ true exactly for $C\in\mathcal{P}\setminus\mathcal{Q}$, and the same calculation shows that it satisfies all constraints; in particular $t\ge0$.

\ref{it:formula-size} Put $A:=(4r)^{2\lambda}$. By \cref{lem:candidates}\ref{it:cand-number} there are at most $A\cdot n$ candidates. They can be listed, with their shares, by going through the sequences described in the proof of that lemma for every vertex $v$ and keeping those that describe a candidate; this takes time polynomial in $A$, $\lambda$, $n$, and $\tot W$, and these numbers are at most $c_r\cdot(n+\tot W+1)$. The formula is then written down in time polynomial in its size. We count its clauses. The partition constraint of a vertex concerns at most $A$ variables and has at most $4A+2\le6A$ clauses by \cref{lem:exactly-one}, so the partition constraints have at most $6A\cdot n$ clauses in total. The list for the budget has at most $A\cdot n\cdot\lambda^2/2$ entries, since every candidate has share at most $\lambda^2/2$, so the budget part has at most $\max\{1,c_0\cdot A\cdot n\cdot\lambda^2/2\}$ clauses by \cref{lem:threshold}. If $n=0$, the formula thus has at most one clause, and $1\le c_r/(16r)$. If $n\ge1$, it has at most
\[
 6A\cdot n+c_0\cdot A\cdot n\cdot\frac{\lambda^2}{2}+1\le c_0\cdot\lambda^2\cdot A\cdot n=\frac{c_r\cdot n}{16r}
\]
clauses, where the inequality uses $7\le\lambda^2/2$ and $c_0\ge1$, and the equality is the definition \cref{eq:cr}.
\end{proof}

\paragraph{The algorithm.}
Let $T$ be a nondecreasing function such that \ThreeSAT{} can be decided in time $T(m)$ (\cref{sec:model}), and fix an algorithm that does so. The algorithm \Sparse{} takes an integer $r\ge1$ and an $r$-sparse instance $(W,k)$ with vertex set $V$; put $n:=|V|$. It proceeds as follows.
\begin{compactenum}[1.]
\item If $n=0$, answer yes.
\item For every set $U\subseteq V$ with $|U|\le8k/\lambda$ and every set $F$ of edges of $G_W[U]$: let $\mathcal{Q}$ consist of the vertex sets of the connected components of the graph with vertex set $U$ and edge set $F$, construct $\varphi_{U,\mathcal{Q}}$, and answer yes if the fixed algorithm for \ThreeSAT{} finds it satisfiable.
\item Answer no.
\end{compactenum}
\begin{proposition}[Sparse instances]\label{prop:sparse}
Let $T$ be a nondecreasing function such that \ThreeSAT{} can be decided in time $T(m)$. Given an integer $r\ge1$ and an $r$-sparse instance $(W,k)$, the algorithm \Sparse{} decides whether $\OPT(W)\le k$ in time
\[
 2^{k/r}\cdot T(c_r\cdot k)\cdot(c_r\cdot(k+\tot W+1))^{O(1)},
\]
where $c_r$ is defined in \cref{eq:cr} and the exponent $O(1)$ does not depend on $r$.
\end{proposition}
\begin{proof}
\emph{Correctness.} If $n=0$, then $\OPT(W)=0\le k$, so Step~1 is correct. If Step~2 answers yes, then $\OPT(W)\le k$ by \cref{lem:formula}\ref{it:formula-sat}. Conversely, suppose that $\OPT(W)\le k$. By \cref{lem:components}\ref{it:comp-connected} there is a partition $\mathcal{P}$ with $\cost_W(\mathcal{P})\le k$ whose clusters induce connected subgraphs of $G_W$. Let $\mathcal{Q}$ be the set of its clusters that are not candidates and $U$ the union of these clusters. Then $|U|\le8k/\lambda$ by \cref{lem:candidates}\ref{it:cand-rest}. Let $F$ be the union of the edge sets of spanning trees of the graphs $G_W[C]$, $C\in\mathcal{Q}$; the connected components of the graph with vertex set $U$ and edge set $F$ have the vertex sets $C\in\mathcal{Q}$. So Step~2 considers the guess $(U,\mathcal{Q})$, and $\varphi_{U,\mathcal{Q}}$ is satisfiable by \cref{lem:formula}\ref{it:formula-sat}.

\emph{Number of guesses.} Step~2 is reached only if $n\ge1$, and then $k\ge1$, because $n\le16rk$ by \cref{def:sparse}. We first count the sets $U$ and then, for each of them, the sets $F$.

Write $N:=16rk$, so $n\le N$. The sets $U$ are the sets of at most $a:=8k/\lambda$ vertices. There are $\binom nu\le\binom Nu$ sets of exactly $u$ vertices, so the number of sets $U$ is at most $\sum_{u\le a}\binom Nu$, where $u$ ranges over the integers with $0\le u\le a$. We claim that
\begin{align}
 \sum_{u\le a}\binom Nu\le\Bigl(\frac{eN}a\Bigr)^a.\label{eq:binomial-sum}
\end{align}
To see this, put $x:=a/N$. Then $0<x\le1$, because $k\ge1$ and $8/\lambda\le1\le16r$. Hence $x^u\ge x^a$ for every $u\le a$, and therefore
\[
 x^a\cdot\sum_{u\le a}\binom Nu\le\sum_{u\le a}\binom Nux^u\le\sum_{u=0}^N\binom Nux^u=(1+x)^N.
\]
The last step is the binomial theorem. Moreover, $(1+x)^N\le e^{xN}=e^a$ since $1+x\le e^x$. Dividing by $x^a=(a/N)^a$ gives \cref{eq:binomial-sum}. Finally, $eN/a=e\cdot16rk\cdot\lambda/(8k)=2er\lambda$, so there are at most $(2er\lambda)^{8k/\lambda}$ sets $U$.

Now fix a set $U$. The graph $G_W[U]$ has at most $2r\cdot|U|$ edges, because every vertex has fewer than $4r$ neighbours, so there are at most $2^{2ra}=2^{16rk/\lambda}$ sets $F$. Hence the number of guesses is at most
\[
 2^{(8k/\lambda)\cdot\log(2er\lambda)}\cdot2^{16rk/\lambda}=2^{k\cdot(16r+8\log(2er\lambda))/\lambda}\le2^{k/r}.
\]
For the last inequality, insert $\lambda=256r^2$: we have $16r/\lambda=1/(16r)$, and $8\log(2er\lambda)/\lambda=\log(512er^3)/(32r^2)\le15/(16r)$ because $\log(512e)<11$ and $\log r\le r$ give $\log(512er^3)<11+3r\le30r$.

\emph{Time per guess.} Let $k\ge1$ and $1\le n\le16rk$ as above. By \cref{lem:formula}\ref{it:formula-size}, the formula of a guess has at most $c_r\cdot n/(16r)\le c_r\cdot k$ clauses and is constructed in time $(c_r\cdot(n+\tot W+1))^{O(1)}$. Since $T$ is nondecreasing, the fixed algorithm for \ThreeSAT{} decides its satisfiability in time at most $T(c_r\cdot k)$. Listing the guesses takes time polynomial in $n$ per guess. Since $T\ge1$ and $n\le16rk\le c_r\cdot k$, each guess thus takes time at most $T(c_r\cdot k)\cdot(c_r\cdot(k+\tot W+1))^{O(1)}$. Multiplying by the number of guesses gives the claimed running time; Step~1 takes constant time.
\end{proof}
 \section{Proofs of the main results}\label{sec:proof}
Phase one (\cref{prop:phase-one}) and the two steps of phase two (\cref{prop:to-sparse,prop:sparse}) are stated for the weighted problem and with separate running times. This section combines them into the results announced in \cref{sec:overview}. The sparsification theorem (\cref{thm:sparsification}) is phase one followed by the reduction to sparse instances; the transfer theorem (\cref{thm:main}) adds the algorithm \Sparse{}; and \cref{cor:main} follows by inserting the known running time for \ThreeSAT{} (\cref{thm:sat}) and letting the threshold $r$ grow slowly with $k$. In each step the work lies in comparing running times, for which we need one calculation in advance.

\subsection{Sparsification}\label{sec:proof-sparsification}
The dense components cost a factor $2^{O(\sqrt k\log k)}$ (\cref{prop:dense}), whereas the search tree of phase one has $2^{(4k\log r)/r}$ leaves. The following inequality shows that the first factor is dominated by a power of the second as long as $r\le\sqrt k$.
\begin{lemma}\label{lem:dense-absorbed}
For all integers $k$ and $r$ with $2\le r\le\sqrt k$ we have $\sqrt k\cdot\log k\le(4k\log r)/r$.
\end{lemma}
\begin{proof}
Put $g(x):=(\log x)/x$ for real $x\ge2$, and $y:=\sqrt k\ge2$. The claim is $g(r)\ge g(y)/2$, since $g(y)/2=(\log k)/(4\sqrt k)$. The derivative of $(\ln x)/x$ is $(1-\ln x)/x^2$, which is negative for $x>e$; so $g$ is decreasing on the real numbers $x\ge4$. Moreover $g(2)=g(4)=1/2$ and $g(3)>1/2$, the latter because $3^2>2^3$ gives $\log3>3/2$. If $r\ge4$, then $g(r)\ge g(y)$ because $4\le r\le y$. If $r\le3$ and $y\ge4$, then $g(r)\ge1/2=g(4)\ge g(y)$. If $r\le3$ and $y<4$, then $g(y)\le(\log4)/2=1\le2\cdot g(r)$.
\end{proof}

We can now prove the sparsification theorem, which we restate from \cref{sec:overview}.
\sparsificationthm*
\begin{proof}
Let $G$ be a graph with $n$ vertices and let $k$ and $r$ be integers with $2\le r\le\sqrt k$. Recall from \cref{sec:model} that $(G,k)$ is a yes-instance of \ClusterEditing{} if and only if $\OPT(W_G)\le k$, that $W_G$ is all-odd, and that $\tot{W_G}\le n^2$.

\emph{The algorithm.} If $k\ge n^2$, then $(G,k)$ is a yes-instance, because deleting all edges takes fewer than $n^2$ edits; output the single pair that consists of the weighted instance without vertices and the budget $0$. This pair is $r$-sparse (\cref{def:sparse}) and has optimum $0$. So assume $k<n^2$. Apply the algorithm of \cref{prop:phase-one} to $r$, $W_G$, and $k$. It outputs at most $2^{(4k\log r)/r}$ instances $W_j$ with budgets $k_j$, each $r$-uniform with at most $n$ vertices, total weight at most $n^2$, and $0\le k_j\le k$, such that $\OPT(W_G)\le k$ if and only if $\OPT(W_j)\le k_j$ for some $j$. Apply the algorithm of \cref{prop:to-sparse} to $r$, $W_j$, and $k_j$ for every $j$. If it reports that $\OPT(W_j)>k_j$, discard $W_j$. Otherwise it outputs an $r$-sparse instance $(W'_j,k'_j)$ with $k'_j\le k_j\le k$ and total weight at most $n^2$ such that $\OPT(W_j)\le k_j$ if and only if $\OPT(W'_j)\le k'_j$; output this instance. Then $(G,k)$ is a yes-instance if and only if $\OPT(W'_j)\le k'_j$ for one of the output instances, and there are at most $2^{(4k\log r)/r}$ of them.

\emph{Running time.} Let $k<n^2$. By \cref{prop:phase-one}, phase one takes time $2^{(4k\log r)/r}\cdot(n+k+n^2)^{O(1)}=2^{(4k\log r)/r}\cdot n^{O(1)}$. By \cref{prop:to-sparse}, the reduction to a sparse instance takes time $(4k+16r)^{4\sqrt k+16r+1}\cdot(n+n^2)^{O(1)}$ for each $j$, because $k_j\le k$. Since $2\le r\le\sqrt k$, the first factor is $2^{O(\sqrt k\log k)}$ by \cref{prop:dense}, and this is $2^{O((k\log r)/r)}$ by \cref{lem:dense-absorbed}. Hence the total running time is $2^{O((k\log r)/r)}\cdot n^{O(1)}$, and none of the hidden constants depends on~$r$.
\end{proof}

\subsection{The transfer theorem}\label{sec:proof-transfer}
\Cref{thm:sparsification} produces $r$-sparse instances, and the algorithm \Sparse{} of \cref{prop:sparse} decides them with the help of an algorithm for \ThreeSAT{}. Its running time contains, besides the calls of that algorithm, a factor polynomial in the number $c_r$ of \cref{eq:cr}, which comes from writing down the formulas. The assumption $r\le k^{1/3}$ of \cref{thm:main} serves to absorb this factor.
We restate the theorem from \cref{sec:overview}.
\transferthm*
\begin{proof}
Let $T$ be a nondecreasing function such that \ThreeSAT{} can be decided in time $T(m)$. Let $G$ be a graph with $n$ vertices and let $k$ and $r$ be integers with $2\le r\le k^{1/3}$; then also $r\le\sqrt k$. Run the algorithm of \cref{thm:sparsification} on $G$, $k$, and $r$, run the algorithm \Sparse{} of \cref{prop:sparse} on $r$ and each of the $r$-sparse instances $(W_i,k_i)$ that it outputs, and answer yes if and only if \Sparse{} answers yes at least once. This is correct by \cref{thm:sparsification,prop:sparse}.

If $k\ge n^2$, the only output instance has no vertices and \Sparse{} answers at once. Let $k<n^2$. The sparsification takes time $2^{O((k\log r)/r)}\cdot n^{O(1)}$ and outputs at most $2^{(4k\log r)/r}$ instances, each with $k_i\le k$ and $\tot{W_i}\le n^2$. By \cref{prop:sparse} and since $T$ is nondecreasing, \Sparse{} takes time at most
\[
 2^{k/r}\cdot T(c_r\cdot k)\cdot(c_r\cdot(k+n^2+1))^{O(1)}\le2^{k/r}\cdot T(c_r\cdot k)\cdot c_r^{O(1)}\cdot n^{O(1)}
\]
on each of them. Here $2^{k/r}\le2^{(k\log r)/r}$ because $\log r\ge1$, and $c_r^{O(1)}=2^{O(r^2\log r)}$ by \cref{eq:cr-growth}, which is $2^{O((k\log r)/r)}$ because $r^3\le k$. Since $T(c_r\cdot k)\ge1$, the total running time is $2^{O((k\log r)/r)}\cdot T(c_r\cdot k)\cdot n^{O(1)}$, and none of the hidden constants depends on $r$ or on $T$.
\end{proof}

\subsection{The running time for \texorpdfstring{\ClusterEditing}{Cluster Editing}}\label{sec:proof-cor}
It remains to insert the bound of \cref{thm:sat} into \cref{thm:main} and to choose the threshold~$r$. The two factors of \cref{thm:main} pull in opposite directions: the factor $2^{O((k\log r)/r)}$ of the search tree shrinks as $r$ grows, whereas the formulas have $c_r\cdot k$ clauses and $c_r$ grows like $2^{O(r^2\log r)}$. The saving of \cref{thm:sat} over exhaustive search is a factor $\sqrt{\log m}$ in the exponent, so the formulas remain affordable as long as $c_r$ is a small power of $\log k$. This allows $r$ to be about $\sqrt{(\log\log k)/(\log\log\log k)}$.
We restate the corollary from \cref{sec:overview}.
\maincor*
\begin{proof}
\emph{The function $T$.} By \cref{thm:sat} there are an integer $a\ge1$ and an algorithm that decides the satisfiability of a 3-CNF formula with $m\ge2$ clauses in time at most $2^{a\cdot m/\sqrt{\log m}}$. A formula with at most one clause is satisfiable unless it consists of the empty clause. Hence \ThreeSAT{} can be decided in time
\[
 T(m):=2^{a\cdot m'/\sqrt{\log m'}}\cdot d,\qquad\text{where }m':=\max\{m,2\},
\]
for a suitable constant $d\ge1$ that accounts for formulas with at most one clause. This function is nondecreasing: for every integer $m\ge2$ we have $\log(m+1)/\log m=1+\log(1+1/m)/\log m\le1+2/m\le(1+1/m)^2$, where we used $\log(1+x)\le x/\ln2\le2x$ and $\log m\ge1$; taking square roots gives $m/\sqrt{\log m}\le(m+1)/\sqrt{\log(m+1)}$. So \cref{thm:main} applies with this $T$.

\emph{The choice of $r$.} By \cref{eq:cr-growth} there is an integer $\gamma\ge1$ with $c_r\le2^{\gamma r^2\log r}$ for every integer $r\ge2$. For an integer $k\ge2^{16}$ put
\[
 \ell:=\log\log k\qquad\text{and}\qquad r:=\Bigl\lfloor\sqrt{\frac{\ell}{4\gamma\log\ell}}\Bigr\rfloor;
\]
note that $\ell\ge4$ and $\log\ell\ge2$. Let $k_0\ge2^{16}$ be an integer such that $\ell/(4\gamma\log\ell)\ge4$ for all $k\ge k_0$; it exists because $\ell/\log\ell$ tends to infinity with $k$, and it depends only on $\gamma$. Let $k\ge k_0$. Then $r\ge2$, and since $\lfloor x\rfloor\ge x/2$ for every real $x\ge1$,
\begin{align}
 \frac12\cdot\sqrt{\frac{\ell}{4\gamma\log\ell}}\le r\le\sqrt{\ell}.\label{eq:r-range}
\end{align}
In particular $r\le\ell\le\log k$, so $r^3\le(\log k)^3\le k$, where the last step uses $x^3\le2^x$ for $x=\log k\ge10$. Hence $2\le r\le k^{1/3}$, as \cref{thm:main} requires.

\emph{The formulas.} By the definition of $r$ we have $r^2\le\ell/(4\gamma\log\ell)$, and $\log r\le\log\ell$ by \cref{eq:r-range}. Hence $\gamma r^2\log r\le\ell/4$ and
\[
 c_r\le2^{\ell/4}=(\log k)^{1/4}.
\]
Since $c_r\cdot k\ge2$ and $\log(c_r\cdot k)\ge\log k$, this gives
\begin{align*}
 T(c_r\cdot k)=d\cdot2^{a\cdot c_r\cdot k/\sqrt{\log(c_r\cdot k)}}&\le d\cdot2^{a\cdot k\cdot(\log k)^{1/4}/\sqrt{\log k}}=d\cdot2^{a\cdot k/(\log k)^{1/4}}\\
 &=d\cdot2^{a\cdot k/2^{\ell/4}}\le d\cdot2^{a\cdot k/\sqrt\ell}.
\end{align*}
The last step uses $2^{\ell/4}\ge\sqrt\ell$, which is equivalent to $\ell\ge2\log\ell$ and holds for $\ell\ge4$.

\emph{The search tree.} By \cref{eq:r-range}, $1/r\le4\sqrt\gamma\cdot\sqrt{(\log\ell)/\ell}$ and $\log r\le\log\ell$, so
\[
 \frac{k\log r}{r}\le4\sqrt\gamma\cdot k\cdot\frac{(\log\ell)^{3/2}}{\sqrt\ell}=4\sqrt\gamma\cdot\frac{k}{s(k)},
\]
with $s(k)=\sqrt\ell/(\log\ell)^{3/2}$ as in \cref{cor:main}.

\emph{Conclusion.} For $k\ge k_0$, run the algorithm of \cref{thm:main} with the function $T$ and the threshold $r$ defined above. By the two estimates, and since $1/\sqrt\ell\le1/s(k)$, it takes time $2^{O(k/s(k))}\cdot n^{O(1)}$. For $k<k_0$, try all sets of at most $k$ pairs of vertices as the set of edits; there are at most $(n^2+1)^{k_0}=n^{O(1)}$ such sets if $n\ge2$. This proves the first statement. Since $s(k)$ tends to infinity with $k$, the running time is $2^{o(k)}\cdot n^{O(1)}$, and therefore at most $c^k\cdot n^{O(1)}$ for every fixed $c>1$.
\end{proof}

 \section{Conclusion}\label{sec:conclusion}
We have shown that a running time for \ThreeSAT{} transfers to \ClusterEditing{} parameterized by the number $k$ of edits (\cref{thm:main}), and that \ClusterEditing{} can therefore be solved in time $2^{o(k)}\cdot n^{O(1)}$ (\cref{cor:main}), as Cao and Chen conjectured~\cite{cao-chen}. The refutation of ETH has removed the lower bounds for many other problems without providing algorithms for them, and our proof raises several questions.

\paragraph{The saving.}
\Cref{cor:main} saves a factor of about $\sqrt{\log\log k}$ in the exponent, whereas \cref{thm:sat} saves $\sqrt{\log m}$. The loss comes from the number $c_r=2^{O(r^2\log r)}$ of \cref{eq:cr}: an $r$-sparse instance allows up to $(4r)^{2\lambda}$ candidate clusters around each vertex, where $\lambda=256r^2$, and our formulas have one variable for each of them. Is there an encoding of $r$-sparse instances whose size is polynomial in $r$? It would improve the saving in \cref{cor:main} to a power of $\log k$. More generally, the reduction of Fomin et al.~\cite[Lemma~18]{fomin-pclusters} shows that \ClusterEditing{} is at least as hard as \ThreeSAT{} with the same saving; is it exactly as hard?

\paragraph{The number of vertices.}
The reduction of Fomin et al.\ also produces graphs with $O(m)$ vertices~\cite[Lemma~18]{fomin-pclusters}, so under ETH there was no algorithm for \ClusterEditing{} with running time $2^{o(n)}$ either. Our algorithm does not provide one, because $k$ can be quadratic in $n$. Can \ClusterEditing{} be solved in time $2^{o(n)}$?

\paragraph{Other edge modification problems.}
\ClusterEditing{} is the problem of editing a graph into one without an induced path on three vertices. Under ETH, most problems of this kind, in which edges are deleted, added, or edited to destroy all induced copies of a fixed graph $H$, have no algorithm with running time $2^{o(k)}\cdot n^{O(1)}$~\cite{aravind-dichotomy,edge-mod-survey}. All these lower bounds have lost their assumption. Which of these problems have subexponential algorithms? The closest relative is Cluster Deletion, in which edges may only be deleted. Our phase one does not apply to it directly: a merge adds up weights of pairs, and a pair that must not be joined has no finite weight.

\paragraph{Solutions that are not sets.}
A solution of \ClusterEditing{} is a partition, and its natural encoding is too long by a logarithmic factor. We removed this factor by first reducing to instances of bounded degree, in which a solution can be assembled from pieces chosen from a family of linear size. The same obstacle arises for problems whose solutions are orderings, such as Feedback Vertex Set, for which no reduction to \ThreeSAT{} that preserves subexponential time is known~\cite{ganian-emso}. Is there a similar route for such problems?
 
\section*{AI use statement}
GPT~6.1 Sol developed the main proof ideas. The manuscript was written in collaboration with Claude Fable~5.2. The authors have carefully checked all claims in the manuscript and take full responsibility for its contents.

\bibliographystyle{plainurl}

\end{document}